\documentclass[prd,onecolumn,showpacs,superscriptaddress,usenatbib]{revtex4}

\def\plottwo#1#2{\centering \leavevmode
\epsfxsize=.34\textwidth \rotatebox{-90}{\epsfbox{#1}} \hfil
\epsfxsize=.34\textwidth \rotatebox{-90}{\epsfbox{#2}}
}

\newcommand{\ThreeJ}[6]{\left (\begin{array}{ccc}#1&#2&#3\\
#4&#5&#6\end{array} \right )}

\newcommand{\SixJ}[6]{\left \{\begin{array}{ccc}#1&#2&#3\\
#4&#5&#6\end{array} \right \} }

\newcommand{\Y}[5]{\: _{#1}Y^{#2}_{#3,#4}(\hat{\mathbf{#5}})}

\newcommand{\blsub}[1]{ \:_{#1} } 

\newcommand{\nhat}[0]{\hat{\mathbf{n}}} 

\newcommand{\mbf}[1]{\mathbf{#1}}
\usepackage{epsfig}
\usepackage{graphicx}
\usepackage{amsmath,amssymb}
\usepackage{aas_macros}     
\newcommand{\cmnt}[1]{{}}
\begin{document}
%
\title{How to detect gravitational waves through the cross-correlation of the galaxy distribution with the CMB polarization}
%
\author{Esfandiar Alizadeh}
\email{ealizad@caltech.edu}

\author{Christopher M. Hirata}
\email{chirata@tapir.caltech.edu}
\affiliation{Caltech M/C 350-17, Pasadena, CA 91125, USA}

\date{January 23, 2012}
%
%
\begin{abstract}
Thompson scattering of cosmic microwave background (CMB) photons off of free electrons during the reionization epoch induces 
a correlation between the distribution of galaxies and the polarization pattern of the CMB, the magnitude of which is proportional to the quadrupole moment of radiation at the time of scattering. Since the quadrupole moment generated by gravitational waves (GWs) gives rise to a different polarization pattern than that produced by scalar modes, one can put interesting constraints on the strength of GWs on large scales by cross-correlating the small scale galaxy distribution and CMB polarization. We use this method together with Fisher analysis to predict how well future surveys can measure the tensor-to-scalar ratio $r$. We find that with a future CMB experiment with detector noise $\Delta_P = 2 \, \mu $K-arcmin 
 and a beam width $\theta_{\rm{FWHM}} = 2 '$ and a future galaxy survey with limiting magnitude $I<25.6$  
 one can measure the tensor-to-scalar ratio with an error $\sigma_r \simeq 0.09$. To measure $r \approx 0.01$, however, one needs $\Delta_P \simeq 0.5 \, \mu $K-radian and $\theta_{\rm{FWHM}} \simeq 1 '$. We also investigate a few systematic effects, none of which turn out to add any biases to our estimators, but they increase the error bars by adding to the cosmic variance. The incomplete sky coverage has the most dramatic effect on our constraints on $r$ for large sky cuts, with a reduction in signal-to-noise smaller than one would expect from the naive estimate $\left( \frac{S}{N} \right)^2 \propto f_{\rm sky}$. Specifically, we find a degradation factor of $f_{\rm deg}=0.32 \pm 0.01$ for a sky cut of $|b|>10^\circ$ ($f_{\rm sky}=0.83$) and $f_{\rm deg}=0.056 \pm 0.004$ for a sky cut of $|b|>20^\circ$ ($f_{\rm sky}=0.66$).
 Nonetheless, given that our method has different systematics than the more conventional method of observing the large scale B modes directly, it may be used as an important check in the case of a detection.  
 
\end{abstract}


\maketitle
%
\section{introduction}
 The possible detection of a primordial B mode signal in the polarization of the Cosmic Microwave Background (CMB) 
will certainly be of great importance for cosmologists and high energy physicists alike as it will provide us with information about physics at high energies that will not be accessible through terrestrial experiments in the foreseeable future (see e.g.~\cite{2009AIPC.1141...10B}). 

 The leading mechanism for setting up the initial condition of the Universe is inflation~\citep{1981PhRvD..23..347G,1982PhLB..108..389L,1982PhRvL..48.1220A}, in which the quantum fluctuations of a scalar field, the inflaton, inside the horizon gets stretched out of the horizon during an almost exponentially expanding phase of the universe, generating the primordial seeds of structure in the Universe~\citep{1981JETPL..33..532M,1982PhLB..115..295H,1982PhLB..117..175S}. The predictions of this theory have so far passed all the existing observational tests (see e.g.~\cite{2011ApJS..192...18K}). However, little is known about the properties of the inflaton field(s) and its potential. 
 A tremendous amount of insight will be gained if the primordial B pattern of the polarization of the CMB can be measured. This pattern can only be generated by primordial {\it tensor} perturbations, that is the gravitational waves (GWs) produced during the inflationary era~\citep{1997PhRvD..55.7368K,1997PhRvD..55.1830Z}. The spectrum of GWs is commonly
 expressed as $\Delta_h(k) = \Delta_h(k_*) (\frac{k}{k_*})^{n_T}$, where $\Delta_h$ is the power spectrum of the traceless-symmetric part of the metric, $h_{ij}$, per logarithmic interval in wavenumber, $k$, and $k_*$ is an arbitrary pivot scale. Furthermore, the strength of the tensor modes is commonly quoted in terms of the ``tensor-to-scalar ratio" $r \equiv \frac{\Delta_h(k_*}{\Delta_{\mathcal R}(k_*)}$, where $\Delta_{\mathcal R}$ is the power per logarithmic interval in wavenumber in the curvature perturbations ${\mathcal R}$. The magnitude of $r$ depends on the Hubble scale during inflation, which is in turn a function of the inflaton potential energy during inflation. Of special importance is the value of $r\sim0.01$ since this value corresponds to a GUT scale energy during inflation so a detection of the signal at this value would strongly suggest a relationship between inflation and GUT scale physics.
 
 However, the signal in the B mode is very small and is peaked on large angular scales where the galactic foregrounds are most difficult to remove. In addition, the lensing of the CMB by large scale structure contaminates the signal by transferring power from the E mode to the B mode \footnote{Although lensing is a source of noise for detecting GWs, it can be used to strengthen the constraints on cosmological parameters that determine the late time properties of the universe, e.g. the equation of state of the dark energy or neutrino masses} . This contaminant can however be removed sufficiently by ''delensing" techniques~\citep{2002ApJ...574..566H,2003PhRvD..67h3002O} to reach the critical accuracy of $r \sim 0.01$~\citep{2002PhRvL..89a1304K,2002PhRvL..89a1303K}. 
 
 Having different methods of measuring the GW signal with different systematic errors is very important to make sure that the detected signal is primordial, not an artifact of instrumental or foreground contaminants such as lensing. The aim of this paper is to propose a new method to measure (or put a bound on) the strength of GWs by using the correlation between the galaxy distribution and the CMB polarization fluctuations. 
That there should be a cross-correlation is clear: Thompson scattering of CMB photons off of free electrons during the reionization era introduces anisotropies to the polarization pattern of the CMB~\citep{1987MNRAS.226..655B}. The amplitude of these anisotropies is proportional to the number density of free electrons, which itself is correlated with the distribution of galaxies, hence the correlation between the galaxy distribution and the CMB polarization. Furthermore, these anisotropies are also proportional to the quadrupole moments of the CMB radiation at the time of scattering. The quadrupole moments that give rise to the B polarization pattern can only be generated by the tensor perturbation of the metric so by measuring the amplitude of the cross-correlation between galaxies and CMB polarization one get an estimate on the amplitude of GW signal.
   
 The outline of the paper is as follows. In section \ref{ss:PfIR} we find an analytical formula for the cross-correlation between the CMB polarization patterns and the galaxy distribution. We will then use this result in section \ref{ss:est} to find a quadratic estimator to construct the electric and magnetic type polarization moments, $\bar{E}^i_{lm}$ and $\bar{B}^i_{lm}$, generated by a smooth electron density field at a given redshift bin $i$ from the observed galaxy distribution and the CMB polarization fluctuations. The details of this calculation can be found in Appendix \ref{app:DotQE}. The noise and signal covariance matrices for detecting these average polarizations are presented in section \ref{ss:SNCM} and the derivation is presented in Appendix~\ref{app:SCM}. In section \ref{sec:StGW} we use the Fisher formalism to forecast the power of this method to constrain the tensor-to-scalar ratio $r$ for futuristic CMB and galaxy surveys. Several possible systematic effects are considered in section \ref{sec:SE}. We conclude in section~\ref{sec:DaC}.
 
 WMAP5 parameters~\cite{2009ApJS..180..330K} are assumed throughout the paper.
%
\section{Formalism} 

\subsection{Polarization from inhomogeneous reionization} \label{ss:PfIR}
We define spin $\pm2$ variables $\blsub{\pm2}P(\nhat)$ on the sphere in terms of the Stokes parameters q and u as~\citep{1997PhRvD..55.1830Z}
\begin{equation}
	\blsub{\pm2}P(\nhat) \equiv [q \pm i u](\nhat).
\end{equation}
These can be found from the integrals of the temperature quadrupole along the line of sight, see e.g. \cite{2000ApJ...529...12H},
\begin{equation}
	\blsub{\pm2}P(\nhat) = \frac{\sqrt{24\pi}}{10} \int dD g(D \nhat)\sum_{m=-2}^2 Q^{(m)}(D \nhat) \Y{\pm2}{}{2}{m}{n},
	\label{equ:Ppm}
\end{equation}
where $D$ is the conformal distance in units of the Hubble distance today, and $g$ is the visibility function.

If we only consider the polarization produced at a given redshift slice ``i," and further separate the visibility function at that slice
 into a smooth part and a part coming from the fluctuations in the electron number density we will have
\begin{equation}
	g^i(D \nhat) = \bar{g}^i(D)\left( 1 + \frac{\delta g^i(D \nhat)}{\bar{g}^i(D)} \right) 
								= \bar{g}^i(D)(1 + \Delta^i_b(D \nhat))
								= \bar{g}^i(D)\left( 1 + \frac{\Delta_g^i(D \nhat)}{b^i} \right).
	\label{equ:g}
\end{equation}
Here, $\Delta_b$ and $\Delta_g$ are the baryon and galaxy overdensities, respectively, and $b$ is the galaxy bias with respect to the baryons.
Similarly, the polarization generated at that redshift slice can be separated into a smooth part and overlying fluctuations. Then
 Eqs.~(\ref{equ:Ppm}) and~(\ref{equ:g}) give
\begin{equation}
	\blsub{\pm}P^i(\nhat) = \blsub{\pm}\bar{P}^i(\nhat) + \delta \blsub{\pm}P^i(\nhat),
\end{equation}
where
\begin{equation}
\blsub{\pm}\bar{P}^i(\nhat) = \frac{\sqrt{24 \pi}}{10} \int dD \bar{g}^i(D) \sum_{m=-2}^{2} Q^{(m)}(D\nhat) \Y{\pm2}{}{2}{m}{n}
\end{equation}
and
\begin{equation}
	\delta \blsub{\pm}P^i(\nhat) \simeq \blsub{\pm}\bar{P}^i(\nhat) \Delta^i(\nhat) \frac{1}{b^i} .
\end{equation}
Here $\Delta^i$ is the projected overdensity of galaxies at that redshift bin
\begin{equation}
	\Delta^i(\nhat) = \frac{1}{D^i_{\rm high}-D^i_{\rm low}} \int_{D^i_{\rm low}}^{D^i_{\rm high}} dD \Delta^i_g(D\nhat) .
\end{equation}

To find the $E$ and $B$ polarization modes coming from that redshift slice we expand all of the functions in terms of their multipole moments
\begin{eqnarray}
	\Delta^i(\nhat) &=& \sum_{l m} \Delta^i_{lm} \Y{}{}{l}{m}{n}, \nonumber \\
	\blsub{\pm}\bar{P}^i(\nhat) &=& \sum_{l m} \blsub{\pm}\bar{P}^i_{lm} \left(\Y{\pm2}{}{l}{m}{n} \right), \nonumber \\
	\delta \blsub{\pm}\bar{P}^i(\nhat) &=& \sum_{l m} \delta \blsub{\pm}\bar{P}^i_{lm} \left(\Y{\pm2}{}{l}{m}{n} \right) .
\end{eqnarray}
We also use the identities
\begin{eqnarray}
	E_{lm} &=& \frac{1}{2}(\blsub{+}P_{lm} + \blsub{-}P_{lm}), \nonumber \\
	B_{lm} &=& \frac{1}{2i}(\blsub{+}P_{lm} - \blsub{-}P_{lm}).  
\end{eqnarray}
Then after a straightforward calculation we find
\begin{eqnarray}
	\delta E^i_{lm} &=& \sum_{l_1 m_1} \sum_{l_2 m_2} (-1)^m \ThreeJ{l}{l_1}{l_2}{-m}{m_1}{m_2} F_{l_1 l l_2} 
										(\alpha_{l_1 l_2 l} \bar{E}^i_{l_1 m_1} - \gamma_{l_1 l_2 l} \bar{B}^i_{l_1 m_1}) \frac{\Delta^i_{l_2 m_2}}{b^i}, \nonumber \\
	\delta B^i_{lm} &=& \sum_{l_1 m_1} \sum_{l_2 m_2} (-1)^m \ThreeJ{l}{l_1}{l_2}{-m}{m_1}{m_2} F_{l_1 l l_2} 
										(\gamma_{l_1 l_2 l} \bar{E}^i_{l_1 m_1} + \alpha_{l_1 l_2 l} \bar{B}^i_{l_1 m_1}) \frac{\Delta^i_{l_2 m_2}}{b^i},	
										\label{equ:deltaEB}								
\end{eqnarray}
where $\alpha$, $\gamma$ and $F$ are defined as
\begin{eqnarray}
	\alpha_{l_1 l_2 l} &=& \frac{1}{2} \left( 1+(-1)^{l_1+l_2+l}\right), \\
\gamma_{l_1 l_2 l} &=& \frac{1}{2i}\left( 1-(-1)^{l_1+l_2+l}\right), \\
	F_{l_1 l l_2} &=& \sqrt{\frac{(2l_1+1)(2l_2+1)(2l+1)}{4\pi}} \ThreeJ{l}{l_1}{l_2}{2}{-2}{0} .
\end{eqnarray}

\subsection{Estimators}  \label{ss:est}
As we can see from the results of the previous section the power in $\delta E_{lm}/\delta B_{l m}$ modes at a given $l$ comes from 
a quadratic sum of a range of multipoles in $\bar{E}_{l_1 m_1}/\bar{B}_{l_1 m_1}$ and $\Delta_{l_2 m_2}$, where $l$, $l_1$ and $l_2$ must 
satisfy the triangle inequalities. Since we are looking for large scale gravitational wave modes, i.e small $l_1$, and because the power in the
galaxy distribution is large at small scales (large $l_2$), we realize that the triangles we are dealing with in $l$ space are elongated
with two long sides of length $l$ and $l_2$ and one small side of length $l_1$. In other words, to find an estimate of the power
of the CMB polarization anisotropies on {\it large} angular scales we can use the {\it small scale} power of the CMB polarization and the galaxy distribution. 

We show in Appendix~\ref{app:DotQE} that an unbiased, minimum variance, quadratic estimator for $\bar{E}^i/\bar{B}^i$ can be found to be
\begin{eqnarray}
	\hat{\bar{E}}^i_{LM} &=& A^{i,E}_L \sum_{l_1 m_1} \sum_{l_2 m_2} (-1)^M \ThreeJ{l_1}{l_2}{L}{m_1}{m_2}{-M} g^{i,E}_{l_1 l_2}(L)
											 \left( \alpha_{l_1 l_2 L} E^{\rm obs}_{l_1 m_1} - \gamma_{l_1 l_2 L} B^{\rm obs}_{l_1 m_1} \right)	
											 \Delta^{i,{\rm obs}}_{l_2 m_2}, 
\label{equ:Eest} \\
	\hat{\bar{B}}^i_{LM} &=& A^{i,B}_L \sum_{l_1 m_1} \sum_{l_2 m_2} (-1)^M \ThreeJ{l_1}{l_2}{L}{m_1}{m_2}{-M} g^{i,B}_{l_1 l_2}(L)
											 \left( \gamma_{l_1 l_2 L} E^{\rm obs}_{l_1 m_1} + \alpha_{l_1 l_2 L} B^{\rm obs}_{l_1 m_1} \right)	
											 \Delta^{i,{\rm obs}}_{l_2 m_2},
\label{equ:Best}											 
\end{eqnarray}
where
\begin{eqnarray}										 
	g^{i,X}_{l_1 l_2}(L)&=& \frac{f^i_{L l_1 l_2}}{M^X_{l_1 l_2 L} C^{g^i g^i,{\rm obs}}_{l_2}}, 
\label{equ:gi}\\
	M^E_{l_1 l_2 L} &=& \left( \left|\alpha_{L l_1 l_2}\right|^2 C^{EE,{\rm obs}}_{l_1}
																						+\left|\gamma_{L l_1 l_2}\right|^2 C^{BB,{\rm obs}}_{l_1}\right), 
\\
	M^B_{l_1 l_2 L} &=& \left( \left|\gamma_{L l_1 l_2}\right|^2 C^{EE,{\rm obs}}_{l_1}
																						+\left|\alpha_{L l_1 l_2}\right|^2 C^{BB,{\rm obs}}_{l_1}\right), 
\\
	f^i_{L l_1 l_2}	&=& F_{L l_1 l_2} \frac{C^{g^i g^i}_{l_2}}{b^i},\label{equ:fi}
\\
	A^{i,X}_L &=& (2L+1) \left[ \sum_{l_1 l_2} f^i_{L l_1 l_2} g^{i,X}_{l_1 l_2}(L)\right]^{-1}	.
\label{equ:Ai}																	
\end{eqnarray}																						
 Here $X$ can be either $E$ or $B$ and $E^{\rm obs}_{lm}$, $B^{\rm obs}_{lm}$ and $\Delta^{\rm obs}_{lm}$ are the observed quantities which can be decomposed as
\begin{eqnarray}
	X^{\rm obs}_{lm}&=& X_{lm} +X^{\rm noise}_{lm} = X^{\rm recom}_{lm} + X^{\rm reion}_{lm} + X^{\rm lens}_{lm} + X^{\rm p.s.}_{lm} + X^{\rm gal}_{lm} + X^{\rm noise}_{lm}, \nonumber \\
	\Delta^{i,{\rm obs}}_{l m} &=& \Delta^i_{lm} + \Delta^{i,{\rm noise}}_{lm},
\end{eqnarray}
where the superscripts are for recombination, reionization, lensing, polarized point sources, polarized galactic foregrounds and noise (detector noise in the case of polarization and Poisson noise for galaxy overdensities). A quantity without any superscript description represents all the sources that contribute to it except the noise. The polarization from the reionization epoch can be separated into the contribution from different redshifts,
\begin{eqnarray}
	X^{\rm reion}_{lm} = \sum_i \left( \bar{X}^i_{lm} + \delta X^i_{lm} \right) ,
\label{equ:Xreion}	
\end{eqnarray}
as are calculated above. The power spectra in the above equations are defined as
\begin{eqnarray}
	\langle \Delta^{i*}_{l m} \Delta^i_{l' m'} \rangle &=& C^{g^i g^i}_l \delta_{l l'} \delta_{m m'}, \nonumber \\
	\langle \Delta^{i,{\rm obs}*}_{l m} \Delta^{i,{\rm obs}}_{l' m'} \rangle &=& C^{g^i g^i,{\rm obs}}_l \delta_{l l'} \delta_{m m'} 
																											 	   =\left( C^{g^i g^i}_l + N^{g^i g^i}_l \right)\delta_{l l'} \delta_{m m'} ,\nonumber \\
	\langle X^{{\rm obs}*}_{l m} X^{{\rm obs}}_{l' m'} \rangle &=& C^{XX,{\rm obs}}_l \delta_{l l'} \delta_{m m'} 
																				= \left( C^{XX}_l + N^{XX}_l \right) \delta_{l l'} \delta_{m m'} .
\label{equ:XXcor}																																
\end{eqnarray}

The noise power spectrum for the galaxy-galaxy correlation is simply the Poisson noise,
\begin{equation}
	N^{g^i g^i}_l = \frac{1}{n^i_{g}},
\end{equation}
where $n^i_g$ is the mean projected number density of galaxies at redshift bin ``i" in units of sr$^{-1}$. 

The instrumental noise power spectrum for either E or B mode polarization can be written as~\cite{1995PhRvD..52.4307K}:
\begin{equation}
	N^{XX}_l = \left(\frac{\Delta_P}{T_{\text{CMB}}}\right)^2 e^{l(l+1)\theta^2_{\text{FWHM}}/8ln2} ,
	\label{equ:CMBnoise}
\end{equation}
where $\Delta_P$ is the detector noise in units of $\mu $K-radian, $T_{\rm CMB} = 2.725 \times 10^6 \, \mu $K 
and $\theta_{\rm FWHM}$ is the width of the beam in units of radians.

\subsection{Signal and Noise covariance matrices for $\hat{\bar{E}}^i_{LM}$ and $\hat{\bar{B}}^i_{LM}$ } \label{ss:SNCM}

The covariance matrices of our estimators for the average polarization generated at a given redshift bin can be written as
\begin{equation}
	\langle \hat{\bar{X}}^{i*}_{LM} \hat{\bar{X}}^{'j}_{L'M'} \rangle = \left(C^{\bar{X}^i \bar{X}^{'j}}_L + N^{\bar{X}^i \bar{X}^{'j}}_L  \right) 
																																				\delta_{L L'} \delta_{M M'},
\end{equation}
where $C^{\bar{X}^i \bar{X}^{'j}}_L$ and $N^{\bar{X}^i \bar{X}^{'j}}_L$ are the signal and the noise covariance matrices, respectively. The noise covariance can be calculated to be
\begin{eqnarray}
	N^{\bar{X}^i \bar{X}^{j}}_L &=& \frac{A^{i,X}_L A^{j,X}_L}{2L+1} \sum_{l_1 l_2} g^{i,X}_{l_1 l_2} g^{j,X}_{l_1 l_2} 
																M^X_{l_1 l_2 L} C^{g^i g^j}_{l_2}.													  
\end{eqnarray}
Since we are looking at the small scale galaxy distribution it is justified to ignore the correlation between the galaxy 
distributions at different redshifts, i.e. $C^{g^i g^j}_l = C^{g^i g^i}_l \delta_{ij}$. The formula for the noise covariance 
matrix then simplifies to
\begin{equation}
	N^{\bar{X}^i \bar{X}^{j}}_L = A^{i,X}_L \delta_{ij}.
\end{equation} 
Parity consideration or a direct calculation shows
\begin{equation}
	N^{\bar{B}^i \bar{E}^j}_L = 0 .
\end{equation}

The signal covariance matrix, $C^{\bar{X}^i \bar{X}^j}_L$, for $E$ and $B$ type polarization can be written as 
\begin{eqnarray}
	C^{\bar{E}^i \bar{E}^j}_L &=& C^{\bar{E}^i \bar{E}^j}_{L,S} + C^{\bar{E}^i \bar{E}^j}_{L,T},
\label{equ:CEEL}\\
	C^{\bar{B}^i \bar{B}^j}_L &=& C^{\bar{B}^i \bar{B}^j}_{L,T},
\end{eqnarray}
where the subscripts $S$ and $T$ mean scalar and tensor, respectively. There is no contribution to the $B$ polarization from 
scalar perturbations. It is worth mentioning that $C^{\bar{E}^i \bar{E}^j}_{L,S}$ actually acts as a source of noise for the purpose of detecting GWs even though it is the main contributer to $\bar{E}^j_{lm}$. 

We show in appendix \ref{app:SCM} that the tensor terms in the above equations can be calculated as
\begin{equation}
	C^{\bar{X}^i \bar{X}^j}_{l,T} = \frac{4}{3\pi} \int dk \: k^2 S^{i,X}_l(k) \: S^{j,X}_l(k) P_h(k),
\end{equation}
where $S^{j,X}_l(k)$ is given in Eq.~(\ref{equ:App_CEiEi}). We follow reference~\cite{2009ApJS..180..330K} in writing 
the power spectrum of GWs as
\begin{equation}
	\Delta_h^2(k) \equiv \frac{k^3 P_h(k)}{2 \pi^2} = \Delta^2_h(k_0)\left( \frac{k}{k_0}\right)^{n_t}
\end{equation} 
for $k_0 = 0.002 \text{ Mpc}^{-1}$. For the purpose of 
this paper we take the tensor spectral tilt to be zero, $n_t = 0$. Furthermore, the amplitude of the GW spectrum can be parametrized in terms of the tensor-to-scaler ratio, r, as
\begin{equation}
	r \equiv \frac{\Delta_h^2(k_0)}{\Delta^2_{\mathcal R}(k_0)} ,
\end{equation}
where $\Delta_{\mathcal R}(k)$ is the curvature perturbation spectrum. We fix $\Delta^2_{\mathcal R}(k_0) = 2.41 \times 10^{-9}$.
 The tensor-to-scalar ratio, r, is then the only parameter in this paper that we try to constrain and the rest of the
 parameters are fixed to their fiducial values.  
 
\begin{figure}
\begin{center}
\includegraphics[width=104mm]{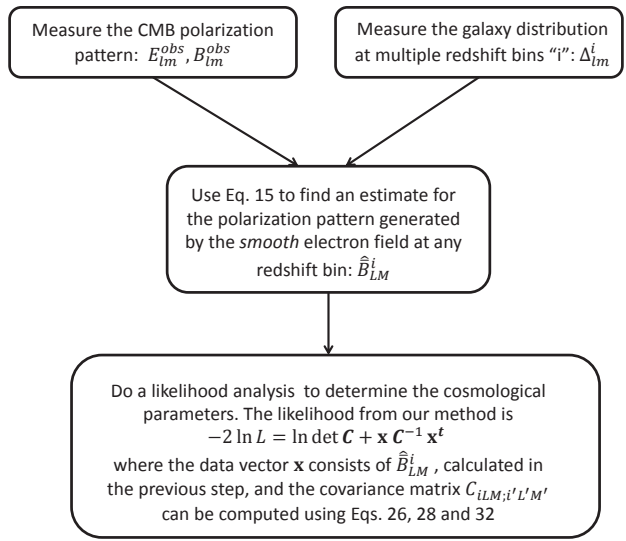}
\end{center}
\caption{ An schematic plot showing how our method can be used as a parameters estimation tool.}
\label{fig:chart}
\end{figure}

 To present the results of this section in a coherent way we show in Fig.~\ref{fig:chart} the steps one needs to take to estimate cosmological parameters using our method. 

%
\section{Sensitivity to gravitational waves} \label{sec:StGW}
In the absence of appropriate data we use the Fisher information method to forecast the capacity of a few futuristic surveys to detect GWs using our method.
  
So far we have found expressions for the covariance matrices of the signal and noise. From them we define
the total covariance matrix as
\begin{eqnarray}
	C_{iLM;i'L'M'} \equiv (\mbf{\lambda}_L)_{ii'} \delta_{LL'} \delta_{MM'},\nonumber \\
	(\mbf{\lambda}_L)_{ii'} \equiv C^{\bar{B}^i \bar{B}^{i'}}_L + N^{\bar{B}^i \bar{B}^{i'}}_L .
\end{eqnarray}
Then assuming that both signal and noise are Gaussian random variables, the Fisher matrix can be written as
\begin{equation}
	F_{a b} = \frac{1}{2} \text{Tr} \left[ \mbf{C}_{,a} \mbf{C}^{-1} \mbf{C}_{,b} \mbf{C}^{-1}\right] ,
\label{equ:Fisher}
\end{equation}
where the derivatives $a$ and $b$ are with respect to the parameters we wish to constrain. The 
inverse of $\mbf{C}$ can be written as
\begin{equation}
	(C^{-1})_{iLM;i'L'M'} = (\mbf{\lambda}^{-1}_L)_{i i'} \delta_{L L'} \delta_{M M'} .
\end{equation}	
Using this in Eq.~(\ref{equ:Fisher}) and performing the sums over the $L$ and $M$ indices we find 
\begin{equation}
	F_{a b} = \sum_{L} \frac{2L+1}{2} \text{Tr} \left[(\mbf{\lambda}_L)_{,a} (\mbf{\lambda}_L^{-1}) 
							(\mbf{\lambda}_L)_{,b} (\mbf{\lambda}_L^{-1}) \right] .
\end{equation}  
The 1-sigma error bars on the parameter $a$ marginalized over the other unknown parameters is $\sqrt{(\mbf{F}^{-1})_{aa}}$, but assuming perfect knowledge of the other parameters this error is $1/\sqrt{(F_{aa})}$. In our analysis we find constraints on the tensor-to-scalar ratio, $r$, by fixing all other 
cosmological parameters to their fiducial values so our Fisher matrix has only one entry and $\sigma_r = 1/\sqrt{F_{rr}}$ .

\begin{figure*}
\plottwo{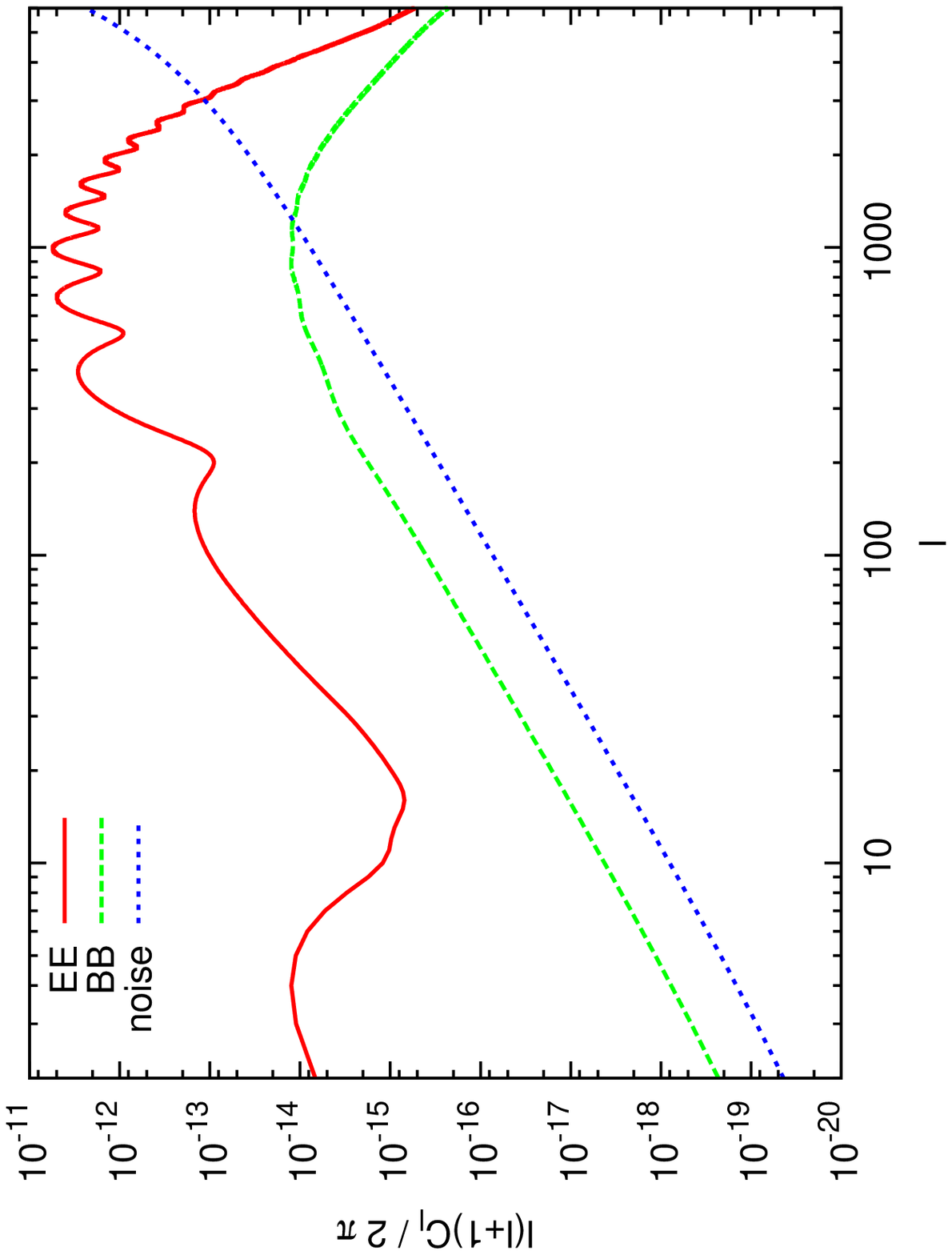}{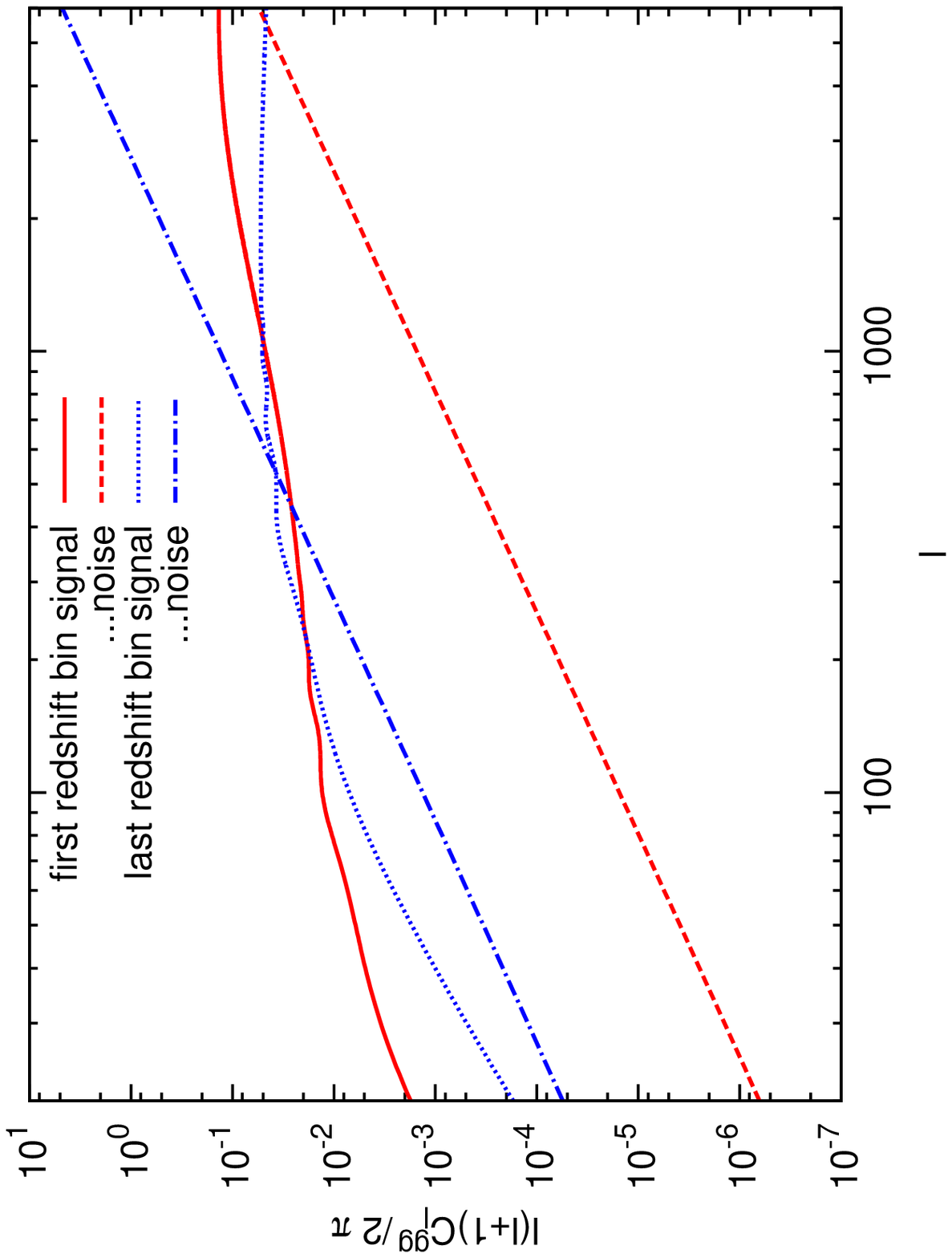}
\caption{[Left] The polarization power spectra for EE (solid red) and BB (dashed green) modes from recombination +
 reionization + lensing as computed by CAMB. The gravitational waves are not included so all the B modes are coming from the lensing of the primordial E modes. The noise power spectrum for a future experiment with 
 $\Delta_p = 5\mu$K-arcmin and $\theta_{\rm FWHM}=2$ arcmin is also shown (dotted blue).
 [Right] Projected galaxy-galaxy signal and noise power spectra for a future survey with limiting magnitude of $I<25.6$, shown for the first and last redshift bins centered at $z=0.5$ and $z=4.1$, respectively, both with a width of $\Delta z =0.2$.}
\label{fig:Cls}
\end{figure*}

To calculate the noise covariance matrix from Eqs.~\ref{equ:gi}--\ref{equ:Ai} we need the polarization power spectra, $C^{XX}_l$, the instrumental noise spectrum, $N^{XX}_l$, the power spectra of the galaxy distribution at different redshifts, $C^{g^i g^i}_l$, their noise spectra $N^{g^i g^i}_l$ 
and the bias $b^i$. We calculate them as follows: 

\noindent We use the publicly available code CAMB~\footnote{http://camb.info/} to calculate $C^{XX}_l$ including lensing. We use Eq.~\ref{equ:CMBnoise} to calculate $N^{XX}_l$ for several choices of detector noise $\Delta_p$ and beam width $\theta_{\text{FWHM}}$. These power spectra are shown in the left panel of figure~\ref{fig:Cls}. The noise power spectrum corresponds to a futuristic experiment with $\Delta_p = 2\, \mu$K-arcmin and $\theta_{\text{FWHM}}=2'$~(for comparison, the detector noise for the ACTPol Deep survey~\cite{2010SPIE.7741E..51N} is $\Delta_P=4 \, \mu$K-arcmin at $\nu =150$ GHz and that of SPTpol survey~\cite{2009AIPC.1185..511M} is $\Delta_p = \sqrt{\frac{2 \Omega_{\rm tot}}{N_{\rm det} t_{\rm obs}}} {\rm NEQ} = 9 \, \mu$K-arcmin at $90$ GHz. These are achieved over a small area but the technology is advancing rapidly and wide sky coverage could be feasible in the near future). The tensor-to-scalar ratio is put to its fiducial value of zero here, so the B mode is exclusively from the weak lensing of the primordial E mode. 

 \noindent The noise in the galaxy power spectrum is $N_l^{g^i g^i} = 1/n^i_{g}$, where $n^i_{g}$ is the mean number of observed galaxies 
 per steradian in the i$^{\rm th}$ redshift bin. To find it we assume a survey similar to that of LSST with limiting magnitude of $I<25.6$ together with a northern counterpart to have a full sky coverage. Our slices in redshift are in the range $0.4<z<4.2$ with a width of $\Delta z = 0.2$. We show the noise power spectrum for the first and last redshift bins in the right panel of Fig.~\ref{fig:Cls} together with the signals at the corresponding redshifts. The signals are calculated from 
 the dark matter power spectrum by using the bias for the star forming galaxies
\begin{equation}
	b(z) = 0.9 + 0.4\:z,
\end{equation} 
which is a fit to the results of Ref.~\cite{2010MNRAS.405.1006O}. We assume baryons trace the dark matter distribution on large scales, 
 so that the same equation for the bias can be used in Eq.~\ref{equ:fi}. 

\begin{figure}
\rotatebox{-90}{\includegraphics[width=3.2in]{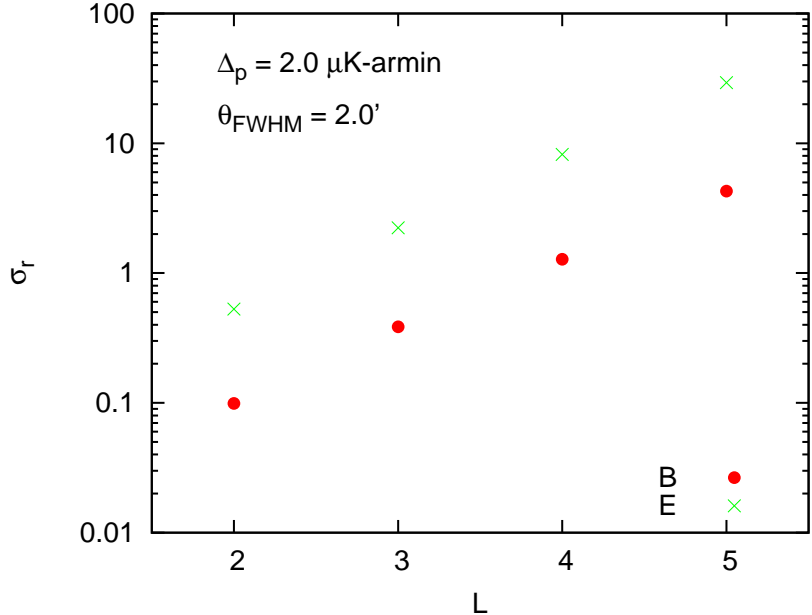}}
\caption{The error bars for measuring the tensor-to-scalar ratio $r$ by constructing either $\hat{\bar{B}}^i_{LM}$ (red circles) 
or $\hat{\bar{E}}^i_{LM}$ (green crosses). The galaxy survey has the specification $I_{\rm max}=25.6$ and $0.4<z<4.2$. The B polarization is more powerful because it is not contaminated by scalar modes. }
\label{fig:error_Ls}
\end{figure}

Given the above specifications we can calculate the one sigma error bars on $r$, $\sigma_r = \sqrt{1/F_{rr}}$. We show in Fig.~\ref{fig:error_Ls} the constraints one gets by constructing either the average magnetic type polarization at each redshift, $\hat{\bar{B}}^i_{LM}$, (red circles) or constructing the electric type polarization, $\hat{\bar{E}}^i_{LM}$, (green crosses) as a function of $L$. As expected, the B modes are more sensitive to the GWs since here the signal is not 
swamped by the scalar modes (see Eq.~\ref{equ:CEEL}). Also, since the signal drops rapidly with increasing $L$ 
(see right panel of Fig.~\ref{fig:Ix_Sk}) most of the information about the GWs is contained in its lowest moment, i.e. the quadrupole. 
 

\begin{figure}
\rotatebox{-90}{\includegraphics[width=3.2in]{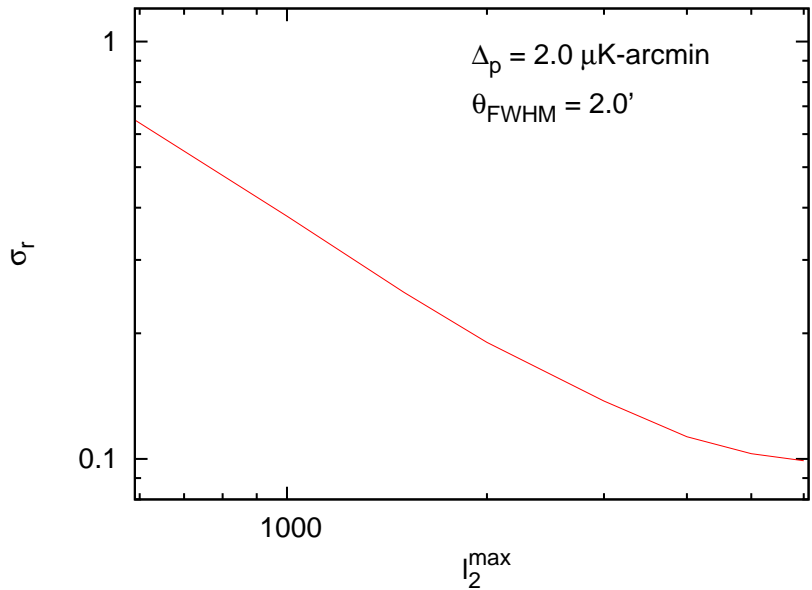}}
\caption{To find out how the signal for detecting $r$ is distributed in multipole space we have plotted the error on measuring the tensor-to-scale ratio versus the maximum $l_2$ in the sum in Eq.~\ref{equ:Ai} . We see that up to the $l_2^{\rm max} \sim 4000$ the signal increases with increasing $l_2^{\rm max}$ after which it levels off.}
\label{fig:converge}
\end{figure}

From Fig.~\ref{fig:Cls} it can be seen that the signal goes below the noise at different values of $l$ for the E or B polarization modes and similarly for the galaxy power spectra at different redshift bins. An interesting question then arises: at what angular scales is most of the signal in determining $\bar{B}_{LM}$ coming from? To find out, we have plotted in Fig.~\ref{fig:converge} the 1-sigma error bars on $r$ as a function of the maximum $l_2$ in the sum in Eq.~\ref{equ:Ai} (the maximum $l_1$ will automatically be determined from the triangle inequality and the value of $L$). We can see that $\sigma_r$ decreases like a power law with increasing $l^{\rm max}_2$ until $l^{\rm max}_2 \sim 4000$ where it starts to level off.

\begin{figure}
\centering
\begin{tabular}{cc}
\epsfig{file=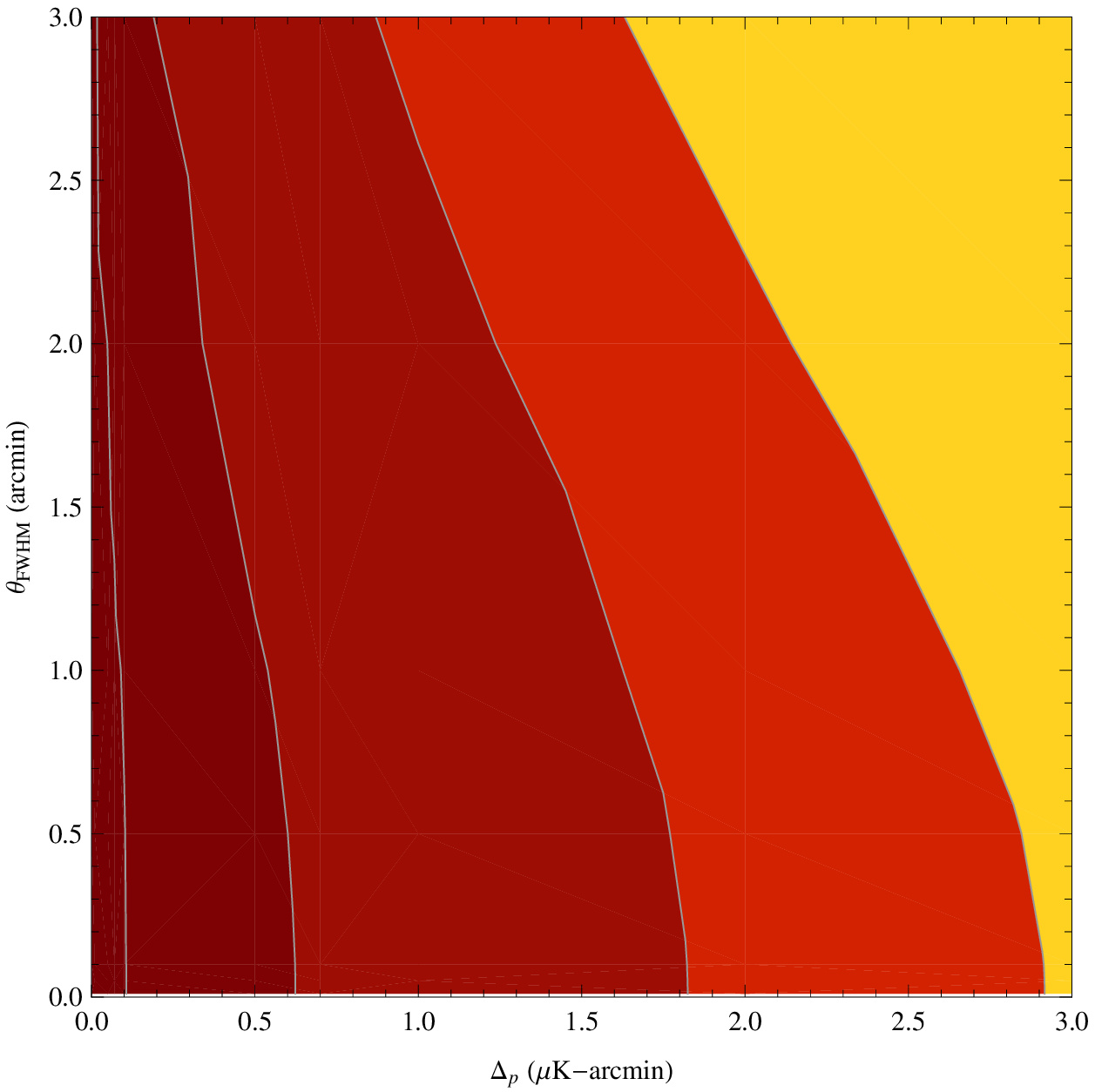,width=0.4\linewidth,clip=} 
\epsfig{file=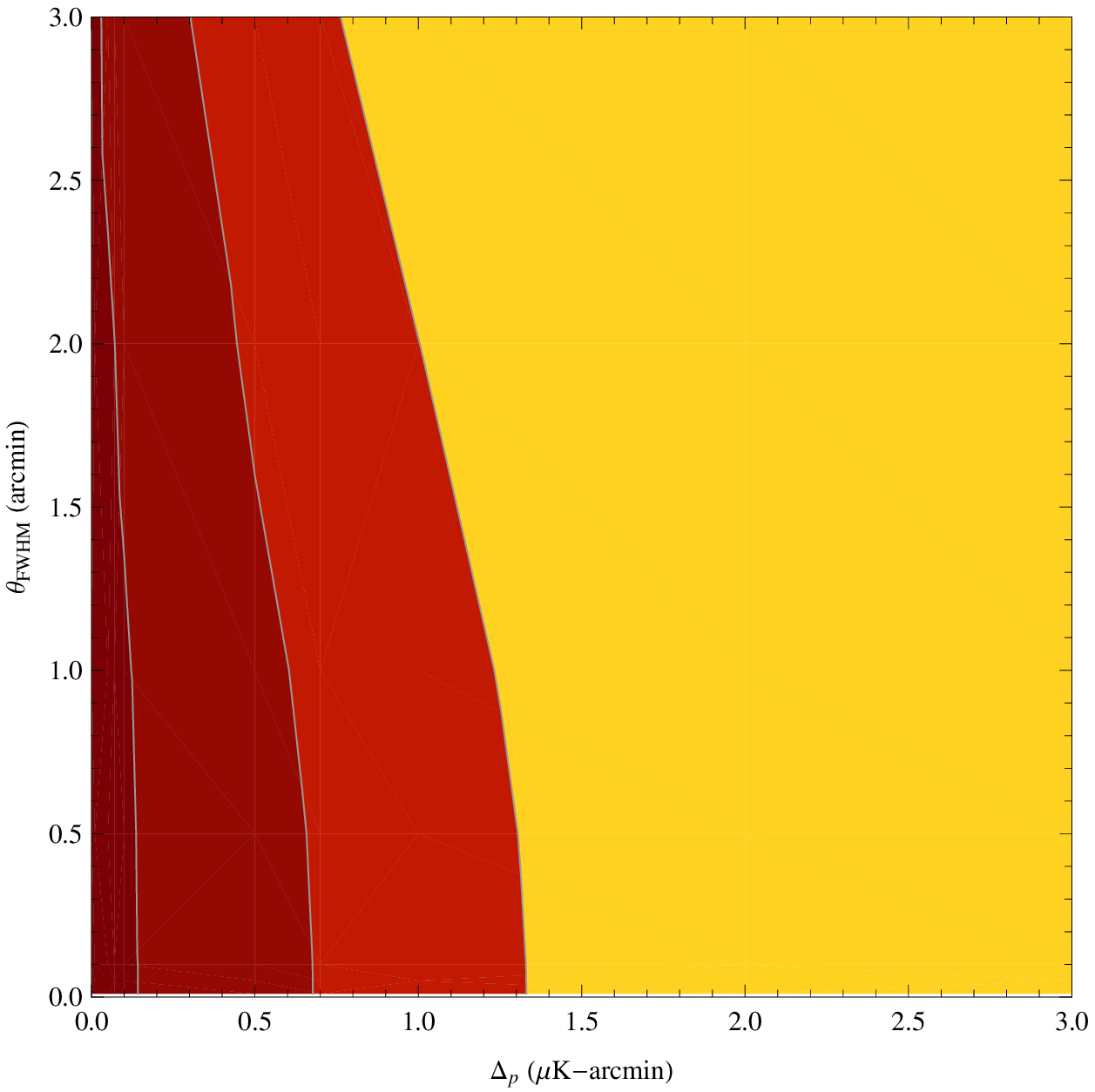,width=0.4\linewidth,clip=} 
\end{tabular}
\caption{Contour plot of the estimated error in the tensor-to-scalar ratio, $\sigma_r$, as a function of detector noise and beam width of a CMB probe. The contours from right to left are for $\sigma_r=0.1$, $0.05$, $0.01$ and $0.001$, respectively. The left panel is for a galaxy survey with maximum limiting magnitude in the $I$ band of $I_{\rm max}=25.6$ and redshift range of $0.4<z<4.2$. The right panel is for a survey capable of observing a constant density of $10^{-3}$ galaxies per ${(\rm Mpc/h)}^3$ from redshift $z=0.4$ up to $z=8$. The latter survey has a smaller number density at low redshifts ($z \lesssim 3.5$) than the former survey but it has a non-negligible surface density of $\sim 10^{6}$ galaxies per square radian up to the highest redshifts. }
\label{fig:contour}
\end{figure}

To explore the dependence of $\sigma_r$ on the specifications of a future CMB and galaxy survey, we show in Fig.~\ref{fig:contour} a contour plot of $\sigma_r$ as a function of detector noise and beam width for two different redshift surveys. The contours are for $\sigma_r=0.1$, $0.05$, $0.01$ and $0.001$ from right to left, respectively. The left panel is for a galaxy survey with $I<25.6$ and $0.4<z<4.2$ and the right panel is for a ``toy" galaxy survey capable of observing $10^{-3}$ galaxies per ${(\rm Mpc/h)}^3$ in the range $4.2<z<8.0$. Comparing these plots we see that the former survey does a better job of constraining $r$ but the difference between these surveys diminishes for smaller values of $\Delta_p$. We conclude that surveys that can observe higher density of galaxies are preferable to the ones that go to the larger redshifts but observe relatively low galaxy surface density. We can also see from these plots that the contours of constant $\sigma_r$ become steep at small $\Delta_p$ so that a small improvement in the detectors sensitivity goes a longer way towards detecting GWs than does decreasing the beam width.

%
\section{Systematic effects} \label{sec:SE}

An understanding of systematic effects is crucial to properly interpret any data set. Biases can be introduced to our parameter estimations either by unaccounted for instrumental errors or by contaminating signals, such as residual foregrounds. One way of finding the extent to which the data has been contaminated by systematic effects is by identifying general properties that the data must satisfy, e.g. invariance under parity, where breakdown of any of these properties is considered a strong indication of an spurious signal. Another way of quantifying the biases is by calculating their effects on parameter estimations, either analytically or by running a suite of Monte-Carlo simulations.

In this section we estimate biases from the gravitational lensing of the CMB, polarized point sources, and incomplete sky coverage. These effects bias our estimators because all of them can introduce correlations between the observed CMB polarization and large scale structure for non-equal values of $l$. This correlation is exactly what we have exploited in our method to estimate the GW signal, so the extra correlation from lensing and point sources, if unaccounted for, can be incorrectly interpreted as a GW signal. It will be shown that these foregrounds will not add any bias to our estimator for $r$ and only increase the error bars by adding to the cosmic variance. In the case of incomplete sky coverage we notice that it is no longer possible to work in multipole space as the E and B modes are non-local functions of the polarization on the sky and therefore cannot be calculated unambiguously. Given that most of our signal comes from very large angular scales ($L=2$) the cuts will have a dramatic effect on our estimates of $r$. Here we set out to calculate these effects quantitatively.

\subsection{Weak lensing of the CMB}

As mentioned before, the observed CMB polarization, $X^{\rm obs}_{lm}$, where $X$ can be either $B$ or $E$, can be written as a sum of the polarization produced around the time of recombination, $X_{lm}^{\rm recom}$, during the reionization era, $X_{lm}^{\rm reion}$, from the gravitational lensing of the primordial polarization, $X_{lm}^{\rm lens}$, from the polarized point sources, $X_{lm}^{\rm p.s.}$, from polarized galactic emission, $X_{lm}^{\rm gal}$, and finally from detector noise $X_{lm}^{\rm noise}$:
\begin{equation}
	X_{lm}^{\rm obs} = X_{lm}^{\rm recom} + X_{lm}^{\rm reion} + X_{lm}^{\rm lens} + X_{lm}^{\rm p.s.} + X_{lm}^{\rm gal} + X_{lm}^{\rm noise} .
\end{equation}

 The polarization produced by the gravitational lensing of the primordial polarization can be written as \citep{2000PhRvD..62d3007H}:
\begin{eqnarray}
	E_{lm}^{\rm lens} =  \sum_{l_5 m_5} \sum_{l_6 m_6} \phi_{l_5 m_5} (-1)^{m_1} \ThreeJ{l_1}{l_6}{l_5}{m_1}{-m_6}{-m_5} ,
				 \blsub{2} F^{\rm lens}_{l_1 l_5 l_6} \alpha_{l_1 l_5 l_6} E^{\rm recom}_{l_6 m_6}, 
\nonumber \\
	B_{lm}^{\rm lens} =  \sum_{l_5 m_5} \sum_{l_6 m_6} \phi_{l_5 m_5} (-1)^{m_1} \ThreeJ{l_1}{l_6}{l_5}{m_1}{-m_6}{-m_5}
				 \blsub{2} F^{\rm lens}_{l_1 l_5 l_6} \gamma_{l_1 l_5 l_6} E^{\rm recom}_{l_6 m_6},		  
\end{eqnarray}
where we have ignored the primordial $B$ mode polarization. Here $\blsub{2} F^{\rm lens}_{l_1 l_5 l_6}$ is defined as
\begin{equation}
	\blsub{2} F^{\rm lens}_{l_1 l_5 l_6} = \left[ l_5(l_5+1) + l_6(l_6+1) - l_1(l_1+1) \right] \sqrt{\frac{(2l_1+1)(2l_5+1)(2l_6+1)}{16 \pi}} \ThreeJ{l_1}{l_5}{l_6}{2}{0}{-2}
\end{equation}
and $\phi$ is the lensing potential, which can be written as a weighted line of sight integral of the gravitational potential, $\Psi$,
\begin{equation}
	\phi(\nhat) = -2 \int_{\eta_{\rm ls}}^{\eta_{0}} d\eta \frac{D_A(\eta-\eta_{\rm ls})}{D_A(\eta_{\rm ls}) D_A(\eta)} \Psi(D_A(\eta)\nhat,\eta) ,
\end{equation}
where $\eta_0$ is the conformal time now and $\eta_{\rm ls}$ that at the last scattering surface.

The bias added to our estimator from the weak lensing of the CMB can then be found by calculating $\left \langle \hat{\bar{B}}^i_{LM} \right\rangle$. A straightforward calculation gives:
\begin{eqnarray}
	\left \langle \hat{\bar{B}}^i_{LM} ({\rm lens}) \right\rangle &=& A^{i,B}_L \sum_{l_1 m_1} \sum_{l_2 m_2} (-1)^M \ThreeJ{l_1}{l_2}{L}{m_1}{m_2}{-M} g^{i,B}_{l_1 l_2}(L)
											 \left( \gamma_{l_1 l_2 L} \left \langle E^{\rm lens}_{l_1 m_1} \Delta^i_{l_2 m_2} \right \rangle + \alpha_{l_1 l_2 L} \left 
											 \langle B^{\rm lens}_{l_1 m_1} \Delta^i_{l_2 m_2} \right \rangle \right) \nonumber \\
											 &=&0 .
\end{eqnarray}
 Therefore, the lensing does not add any biases to our estimator. However, lensing acts as a source of noise by contributing to the background covariance matrix $\left \langle \hat{\bar{B}}^{i*}_{LM}  \hat{\bar{B}}^j_{L'M'} \right\rangle$, i.e. it adds to the cosmic variance. To calculate this noise, we need to find ensemble averages of the form $\langle X^{{\rm obs}*}_{l_1 m_1} \Delta^{i*}_{l_2 m_2}  {X'}^{\rm obs}_{l_3 m_3} \Delta^{j}_{l_4 m_4}\rangle$. Ignoring non-Gaussianities, this ensemble average can be expanded into three terms. The noise coming from the terms of the form $\langle X^{{\rm obs}*}_{l_1 m_1} {X'}^{\rm obs}_{l_3 m_3}  \rangle \langle \Delta^{i*}_{l_2 m_2} \Delta^{j}_{l_4 m_4}\rangle$ have already been taken into account by using the total $C^{EE,{\rm obs}}_{l}$ and $C^{BB,{\rm obs}}_l$, including lensing, in our formulas for the noise covariance matrix (see Eqs.~\ref{equ:gi}-\ref{equ:Ai}). The extra noise coming from the other two contractions is:
\begin{eqnarray}
	&&\langle \hat{\bar{B}}^{i*}_{LM} \hat{\bar{B}}^{j}_{L'M'} \rangle({\rm lens-lens}) = 
			\delta_{L L'} \delta_{M M'} \frac{A^{i,B}_L A^{j,B}_{L}}{2L+1} \Big[ \nonumber \\	
	&& \sum_{l_1} \sum_{l_2} \sum_{l_3} \sum_{l_4} \sum_{l_5} 
			(-1)^{l_1+l_3} g^{i,B}_{l_1 l_2}(L) g^{j,B}_{l_3 l_4}(L) 
			\blsub{2} F^{\rm lens}_{l_1 l_4 l_5} \blsub{2} F^{\rm lens}_{l_3 l_2 l_5} 
			<l_2 l_4 l_5 L> 
		  \SixJ{l_3}{l_4}{L}{l_1}{l_2}{l_5} C^{EE,{\rm recom}}_{l_5}  C^{\phi j}_{l_4} C^{\phi i}_{l_2} \Big] ,
\end{eqnarray}
where $\SixJ{}{}{}{}{}{}$ is the Wigner 6-J symbol, and $<l_2 l_4 l_5 L> =1$ when $l_2+l_4+l_5+L =$odd and zero otherwise. The correlation between the lensing potential and the galaxy distribution at redshift slice $i$ is defined as $\langle \phi^*_{lm} \Delta^i_{l' m'} \rangle = C^{\phi i}_l \delta_{ll'}\delta_{mm'}$. However, this term is subdominant compared to the term we have already considered.

To find out to what extent weak lensing of the CMB contaminates the signal we have computed the expected errors on $r$ assuming no lensing of the CMB. Even though this is an unrealistic assumption it will serve to illustrate the degradation in the signal caused by the gravitational lensing of the CMB. We find that the error on $r$ reduces to $\sigma_r \simeq 0.04$ for complete cleaning of the lensing, compared to $\sigma_r \simeq 0.09$ for no lensing cleaning, so at best the signal can be improved by a factor of $\sim 2$.

\subsection{Polarized Point Sources}
Since the distribution of polarized point sources follows that of dark matter they can add spurious signal to our estimate of $r$, if uncleaned, by correlating the galaxy distribution and the CMB polarization fluctuations. In this section we quantify this effect by calculating the bias and noise introduced to our estimator of the tensor-to-scalar ratio by polarized point sources .

The E/B polarization multipoles for point sources can be found to be:
\begin{eqnarray}
	E^{p.s.}_{lm} \pm i B^{p.s.}_{lm} &=& \int  d^2 \nhat \Y{\pm 2}{*}{l}{m}{n} \left[ Q^{p.s.}(\nhat) \pm i U^{p.s.}(\nhat) \right]
\nonumber \\
	&=&	\int  d^2 \nhat \Y{\pm 2}{*}{l}{m}{n} \left[ \sum_i S_i e^{\pm 2 \omega_i} \delta_D(\nhat - \nhat_i) \right] 
\nonumber \\
	&=&  \sum_i \blsub{\pm 2} Y^*_{l,m}(\nhat_i) S_i 	e^{\pm 2 \omega_i} .
\end{eqnarray}
 Here $S_i$ is the polarized flux of the $i$th point source, $\nhat_i$ is its direction on the sky, and $\omega_i$ is the direction of its polarization with respect to the $\hat{e}_{\theta}$ unit vector on the sphere.
 
 Similar to the lensing case, to find the bias introduced to our estimator from point sources we need to calculate the contribution to our estimator $\left \langle \hat{\bar{B}}^i_{LM} \right \rangle$ from point sources, where now the ensemble average also includes averaging over $S_i$, $\nhat_i$, and $\omega_i$. This average is trivially zero because $\langle e^{\pm 2 \omega_i} \rangle_{\omega_i} = 0$. So the polarized point sources do not add any bias to our estimator.
 
 Calculation of the noise contributed by point sources is a bit more involved. The final result is:
\begin{eqnarray}
	&&\langle \hat{\bar{B}}^{i*}_{LM} \hat{\bar{B}}^{j}_{L'M'} \rangle({\rm p.s.-p.s.}) = 
			\delta_{L L'} \delta_{M M'} \frac{A^{i,B}_L A^{j,B}_{L}}{2L+1} \frac{\langle \sum_{p} S_p^2 \rangle}{4} \Big[ \nonumber \\	
	&& \sum_{l_1} \sum_{l_2} \sum_{l_3} \sum_{l_4} \sum_{l_5} 
			(-1)^{l_3+l_4+l_5} g^{i,B}_{l_1 l_2}(L) g^{j,B}_{l_3 l_4}(L) 
			\gamma_{l_1 l_2 L} \gamma_{l_3 l_4 L} B^{ij}_{l_2 l_5 l_4} \ThreeJ{l_1}{l_3}{l_5}{-2}{2}{0} 
		  \SixJ{l_4}{l_3}{L}{l_1}{l_2}{l_5}  \Big] ,
\end{eqnarray}
where the bispectrum $B$ is defined as:
\begin{eqnarray}
	\langle \Delta^i_{l_2 m_2} \Delta^{\rm p.s.}_{l_5 m_5} \Delta^j_{l_4 m_4} \rangle_{\rm LSS} = \ThreeJ{l_2}{l_5}{l_4}{m_2}{m_5}{m_4} B^{ij}_{l_2 l_5 l_4} .
\end{eqnarray}
Here $\Delta^{\rm p.s.}$ is the projected overdensity of point sources on the sky.
 
 \subsection{Incomplete Sky Coverage}
 
 A real CMB$\times$LSS cross-correlation will have only partial sky coverage, due at the very least to the Galactic Plane, and possibly to additional observing constraints (e.g. hemispherical). This is troublesome for our method because most of the signal comes from large angular scales (see Fig.~\ref{fig:error_Ls}). Even though it is possible to clean these foregrounds by, for example, using the frequency dependence of synchrotron and dust emission, this procedure might introduce unwanted correlations in the residual polarization maps, which in turn can be mistaken for a signal in our estimator. In this paper we take a more approach whereby we completely cut a portion of sky that is most severely contaminated by polarized galactic emission. However, by doing this we can no longer use our estimators, written in multipole space, since the $E_{lm}$ and $B_{lm}$ coefficients are nonlocal functions of the polarization in real space and for a given $l$ they cannot be reliably estimated for all possible $m$. Consequently, our estimators for $\hat{\bar{E}}_{LM}$ $\hat{\bar{B}}_{LM}$ are not applicable in this situation. 
 
 Here we outline the procedure to overcome this issue and refer the interested reader to Appendix~\ref{app:SCu} for details. To solve this problem, one needs to work in configuration space.  We first find an estimator for the average polarization generated at a given redshift bin ``i" coming from a given direction $\bf{n}$ on the sky, $\hat{\bar{\chi}}^i({\bf n})$, where $\chi$ stands for the Stokes parameters $Q$ or $U$. We then compute the $2N_z N_{\rm pix} \times 2N_z N_{\rm pix}$ covariance matrix $C^{ij}_{\chi_A \chi_B} = \left \langle \hat{\bar{\chi}}^{i*}_A({\bf n}_A) \hat{\bar{\chi}}^j_B({\bf n}_B) \right \rangle$ for data consisting of $\bar{Q}^i({\bf n})$ and $\bar{U}^i({\bf n})$ measured in $N_{\rm pix}$ pixels and $N_z$ redshift bins. This covariance matrix consists of a part proportional to $r$ coming from gravitational waves and a part from all other sources. In a Fisher analysis these modes can be projected out by giving them formally infinite power. Finally, we calculate the error bars on $r$ by computing the Fisher matrix using a Monte Carlo code. 
 
We report our results in terms of a ``degradation factor" $f_{\rm deg}$ defined as the ratio of $\sigma_r$ for a cut sky to its value for complete sky coverage. We have computed $f_{\rm deg}$ using the Monte Carlo trace code of Ref.~\cite{2004PhRvD..70j3501H}. Computations were done pixelized at {\sc HEALPix} resolution 8, which has 786$\,$432 pixels \cite{2005ApJ...622..759G}, although convergence was found at resolution 7. We chose for the $E$-mode projection a Gaussian prior $\lambda_\ell = 50e^{-\ell(\ell+1)\sigma^2/2}$ with $\sigma = 0.079\,$rad$\, =  4.5\,$deg; the important point is for the low $\ell$s to have $\lambda_\ell\gg 1$. A total of 200 Monte Carlo trace realizations were run for each sky cut. We find a degradation factor of $f_{\rm deg}=0.32 \pm 0.01$ for a sky cut of $|b|>10^\circ$ ($f_{\rm sky}=0.83$) and $f_{\rm deg}=0.056 \pm 0.004$ for a sky cut of $|b|>20^\circ$ ($f_{\rm sky}=0.66$). Note in particular that $f_{\rm deg}<f_{\rm sky}$ for these cases, and that $f_{\rm deg}\approx f_{\rm sky}$ is a poor approximation.

%
\section{Discussion and Conclusion} \label{sec:DaC}
Part of the polarization pattern of the CMB is generated by the Thompson scattering of photons off of free electrons. The strength of this effect is proportional to the local quadrupole moment of the radiation and the number density of free electrons. The polarization that is generated during the reionization era is therefore modulated on small angular scales by the varying electron number density. This variation in electron number density itself is correlated with the distribution of galaxies. So, by looking for correlations between the small scale CMB polarization fluctuations with the galaxy number density at a given redshift one can determine the local quadrupole moment of the CMB at that redshift. These quadrupoles at different patches of the sky and at different redshifts then provide us with a map of the quadrupole moments during the reionization era~\footnote{In this sense our method is similar to the one proposed in Ref.~\cite{1997PhRvD..56.4511K}. However, those authors suggest the use of a handful of clusters to reconstruct the quadrupole moments at their locations but our method uses a random field of galaxies to reconstruct the quadrupole field. See also Refs.~\cite{2003PhRvD..67f3505C,2004ApJ...612...81D,2004astro.ph..2474S,2004PhRvD..70f3504P,2006PhRvD..73l3517B}}. A small part of this quadrupole pattern can be produced by the tensor modes of fluctuations, i.e. gravitational waves. Therefore, the correlation between galaxy distribution and the CMB polarization anisotropies can be used to constrain the strength of the primordial GWs, which are of the utmost importance for physicists. They provide us with information about the physical processes at work in the early Universe with energies way beyond the reach of any terrestrial experiment. 

In this paper we have investigated the prospect of using this method to measure the tensor-to-scalar ratio $r$, which is a measure of the strength of the GWs. We have constructed a full sky estimator for measuring the average $B$ and $E$ polarization generated at any given redshift bin during the reionization era by using the observed distribution of galaxies at that redshift and the observed  CMB polarization pattern. The result of this exercise was then used, together with linearized perturbation theory and Fisher formalism, to predict the prospects of detecting a GW signal from future experiments. We have found that a CMB experiment with noise parameters of $\Delta_P = 2 \mu$K-arcmin and $\theta_{\rm FWHM} =2$~arcmin, together with a LSST-like galaxy survey~\footnote{http://www.lsst.org} with a limiting magnitude of $I<25.6$ and a similar northern component~\footnote{ A northern component could in principle be done with HyperSuprimeCam/HSC (http://oir.asiaa.sinica.edu.tw/hsc.php) but this is not planned.} can be used to constrain $r$ to $\sigma_r \simeq 0.09$. Even though these are photometric redshift surveys the small fraction of incorrect redshift determination will not be a source of problem for our method because it cannot produce spurious B modes in $\bar P$.

To reach this level of accuracy many obstacles need to be overcome, including the proper handling of the systematic effects. Here, we investigated two of these effects: the polarization induced by the weak gravitational lensing of the CMB and polarized point sources. We found that neither of these contaminants adds a bias to our estimator. However, they increase the error bars on $r$ by increasing the cosmic variance by adding to the observed CMB power $C^{BB,{\rm obs}}_l$ and $C^{EE,{\rm obs}}_l$. We showed that lensing increases $\sigma_r$ by a factor of $\sim 2$, so lensing cleaning techniques might partially help to improve the signal. The most important factor, however, turns out to be the incomplete sky coverage where the degradation in signal to noise is more severe than the naive estimate $\left(\frac{S}{N}\right)^2 \propto f_{\rm sky}$. For example, for a sky cut of $|b|>10^\circ$ ($f_{\rm sky}=0.83$)) the depredation factor is $f_{\rm deg}=0.32$ and for $|b|>20^\circ$ ($f_{\rm sky}=0.66$) it is $f_{\rm deg}=0.056$.
   

\acknowledgements

E.A. and C.H. were supported by the U.S. Department of Energy (DE-FG03-92-ER40701). C.H. is also supported by the National Science Foundation (AST-0807337) and David \& Lucile Packard Foundation. E.A. acknowledges support from the National Science Foundation (grant AST 07-08849) during part of this work. He thanks Laura Book and Ben Wandelt for their help and Avi Loeb and John Kovac for useful conversations.

\appendix
\section{Derivation of the quadratic estimator} \label{app:DotQE}

In this section we derive formula \ref{equ:Eest} for the $\bar{E}^i$ estimator. The derivation for $\bar{B}^i$ is similar. This calculation closely resembles the lensing reconstruction method of Ref.~\cite{2003PhRvD..67h3002O}. 

We start from a general quadratic estimator of the form 
\begin{equation}
	\hat{\bar{E}}^i_{LM} = A^{i,E}_L \sum_{l_1 m_1} \sum_{l_2 m_2} (-1)^M \ThreeJ{l_1}{l_2}{L}{m_1}{m_2}{-M} 
												\left( g^{i,EE}_{l_1 l_2}(L) E^{\rm obs}_{l_1 m_1} + g^{i,EB}_{l_1 l_2}(L) B^{\rm obs}_{l_1 m_1} \right) 
												\Delta^{i,{\rm obs}}_{l_2 m_2}
\end{equation}
and our goal is to find the unknowns $A_L$, $g^{EE}$ and $g^{EB}$. To find them we use two common criteria for a good estimator,
that is that it should be unbiased and have minimum variance.

For the estimator to be unbiased we need 
\begin{equation}
	\langle \hat{\bar{E}}^i_{LM} \rangle \Big|_{\text{small angles}} = \bar{E}^i_{LM},
\end{equation}
that is the ensemble average of the estimator over different realizations of the small angular scales of the polarization and galaxy 
distribution should be equal to its true value. We must then have
\begin{equation}
	 A^{i,E}_L \sum_{l_1 m_1} \sum_{l_2 m_2} (-1)^M \ThreeJ{l_1}{l_2}{L}{m_1}{m_2}{-M} 
												\left( g^{i,EE}_{l_1 l_2}(L) \langle E^{\rm obs}_{l_1 m_1} \Delta^{i,{\rm obs}}_{l_2 m_2} \rangle
												+ g^{i,EB}_{l_1 l_2}(L) \langle B^{\rm obs}_{l_1 m_1} \Delta^{i,{\rm obs}}_{l_2 m_2} \rangle \right) 
												= \bar{E}^i_{LM} .		
\label{equ:A3}																			
\end{equation}
The correlations between the observed galaxy and polarization distributions can be calculated using Eqs.~(\ref{equ:Xreion}-\ref{equ:XXcor})
\begin{eqnarray}
	\langle E^{\rm obs}_{l m} \Delta^{i,{\rm obs}}_{l' m'} \rangle = \langle \delta E^i_{l m} \Delta^i_{l' m'} \rangle
		&=&  \sum_{l_1 m_1} \sum_{l_2 m_2} (-1)^m \ThreeJ{l}{l_1}{l_2}{-m}{m_1}{m_2} F_{l_1 l l_2} 
										(\alpha_{l_1 l_2 l} \bar{E}^i_{l_1 m_1} - \gamma_{l_1 l_2 l} \bar{B}^i_{l_1 m_1})
										\frac{C^{g^i g^i}_{l'}}{b^i} \delta_{l_2,l'} \delta_{m_2,-m'}(-1)^{m'} \nonumber \\
		&=&	\sum_{l_1 m_1} (-1)^{m+m'} \ThreeJ{l}{l_1}{l'}{-m}{m_1}{-m'} F_{l_1 l l'} 
				(\alpha_{l_1 l l'} \bar{E}^i_{l_1 m_1} - \gamma_{l_1 l l'}\bar{B}^i_{l_1 m_1})	\frac{C^{g^i g^i}_{l'}}{b^i} .			
\end{eqnarray}
Using the identity $\ThreeJ{l_1}{l_2}{l_3}{-m_1}{-m_2}{-m_3} = (-1)^{l_1+l_2+l_3} \ThreeJ{l_1}{l_2}{l_3}{m_1}{m_2}{m_3}$, 
the fact that an odd permutation of columns of the 3-j symbol introduces a similar factor of $(-1)^{l_1+l_2+l_3}$, the fact
that for the 3-j symbol to be non-zero we must have $m_1+m_2+m_3=0$, and changing the indices from $(l_1,m_1)$ to $(L,M)$,
 we finally find
\begin{eqnarray}
	&&\langle E^{\rm obs}_{l m} \Delta^{i,{\rm obs}}_{l' m'} \rangle \Big|_{\text{small angles}}  = \sum_{LM}(-1)^M\ThreeJ{l}{l'}{L}{m}{m'}{-M}f^i_{L l l'} 
																		  \left( \alpha_{Lll'} \bar{E}^i_{LM} - \gamma_{L l l'} \bar{B}^i_{LM}\right),
\nonumber \\
	&&f^i_{L l l'} \equiv F_{L l l'}\frac{C^{g^i g^i}_{l'}}{b^i} .		
\end{eqnarray}														   
A similar calculation gives
\begin{eqnarray}
	&&\langle B^{\rm obs}_{l m} \Delta^{i,{\rm obs}}_{l' m'} \rangle \Big|_{\text{small angles}}  = \sum_{LM}(-1)^M\ThreeJ{l}{l'}{L}{m}{m'}{-M}f^i_{L l l'} 
																		  \left( \gamma_{Lll'} \bar{E}^i_{LM} + \alpha_{L l l'} \bar{B}^i_{LM}\right).			
\end{eqnarray}
Plugging these results into Eq.~(\ref{equ:A3}) we find a relation between the unknown coefficients of the form
\begin{eqnarray}
	\frac{g^{i,EE}_{l_1 l_2}(L)}{\alpha_{L l_1 l_2}} = - \frac{g^{i,EB}_{l_1 l_2}(L)}{\gamma_{L l_1 l_2}} \equiv g^{i,E}_{l_1 l_2}(L) , \nonumber \\
	A^{i,E}_L = (2L+1) \left\{ \sum_{l_1 l_2} f^i_{L l_1 l_2} g^{i,E}_{l_1 l_2}(L) \right\}^{-1} .
\end{eqnarray}
Using these relations, the only unknown coefficients are $g^{i,E}_{l_1 l_2}(L)$. To find them we demand that the estimator have
 minimum variance
\begin{eqnarray}
	\frac{ \partial \langle \hat{\bar{E}}^{i,*}_{LM} \hat{\bar{E}}^{i}_{LM} \rangle }{\partial g^{i,E}_{l' l''}(L)}=0 ,
\label{equ:A11}	
\end{eqnarray}
where the variance can be calculated to be
\begin{eqnarray}
	&&\langle \hat{\bar{E}}^{i,*}_{LM} \hat{\bar{E}}^{i}_{LM} \rangle = C^{\bar{E}^i \bar{E}^i}_L + 
				\frac{(A^{i,E})^2}{2L+1} \sum_{l_1 l_2} (g^{i,E}_{l_1 l_2}(L))^2 M^E_{L l_1 l_2} C^{g^i g^i,{\rm obs}}_{l_2} , 
				 \\
	&&			M^E_{l_1 l_2 L} = \left(\left|\alpha_{L l_1 l_2} \right|^2 C^{EE,{\rm obs}}_{l_1} + \left|\gamma_{L l_1 l_2}\right|^2  
					C^{BB,{\rm obs}}_{l_1}\right) .
\end{eqnarray}
Using this in Eq.~(\ref{equ:A11}) we finally obtain
\begin{equation}
	g^{i,E}_{l_1 l_2}(L) = \frac{f^i_{L l_1 l_2}}{M^E_{l_1 l_2 L} C^{g^i g^i,{\rm obs}}_{l_2}} .
\end{equation}

\section{Signal covariance matrix}
\label{app:SCM}
\begin{figure*}
\plottwo{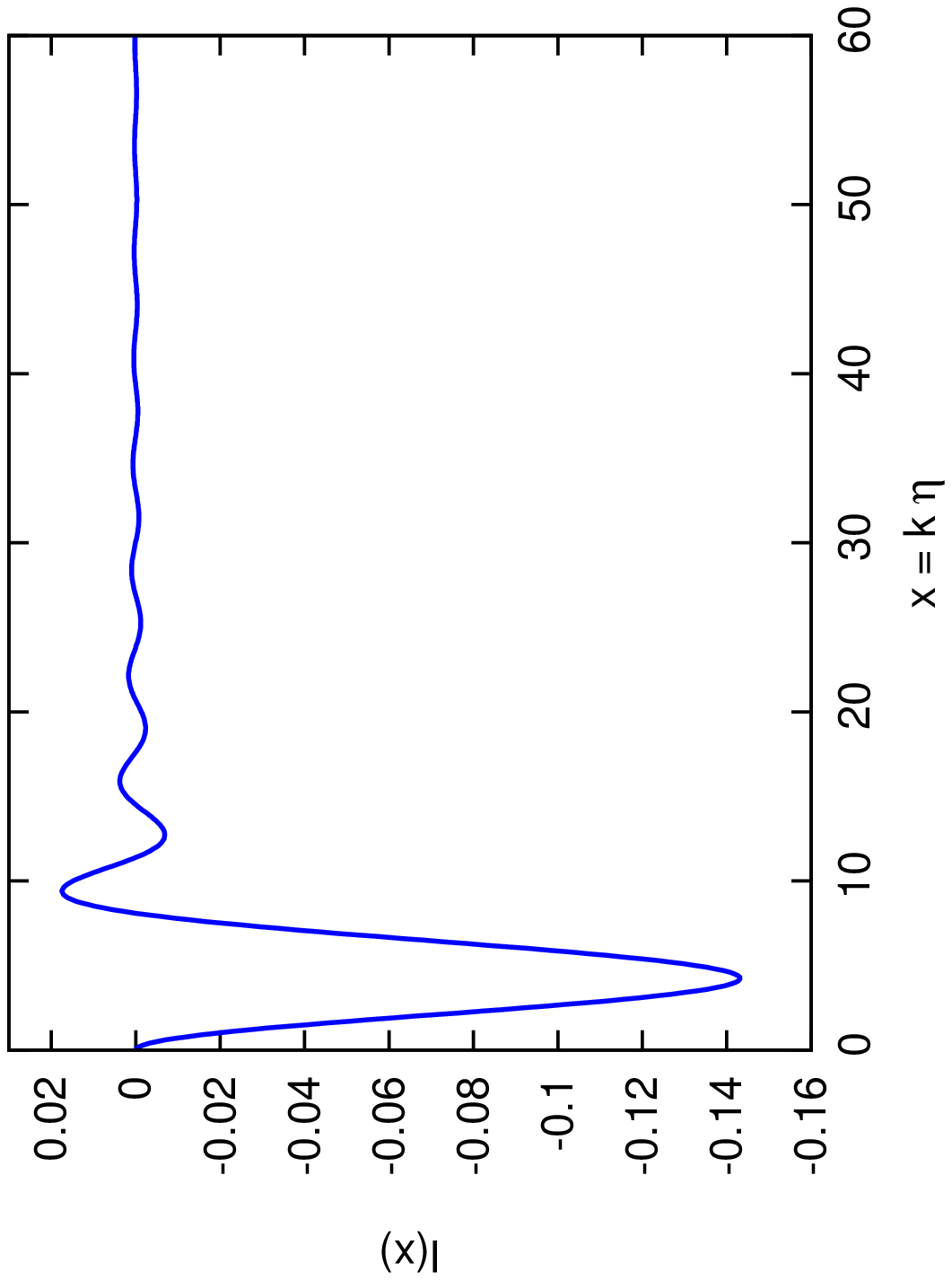}{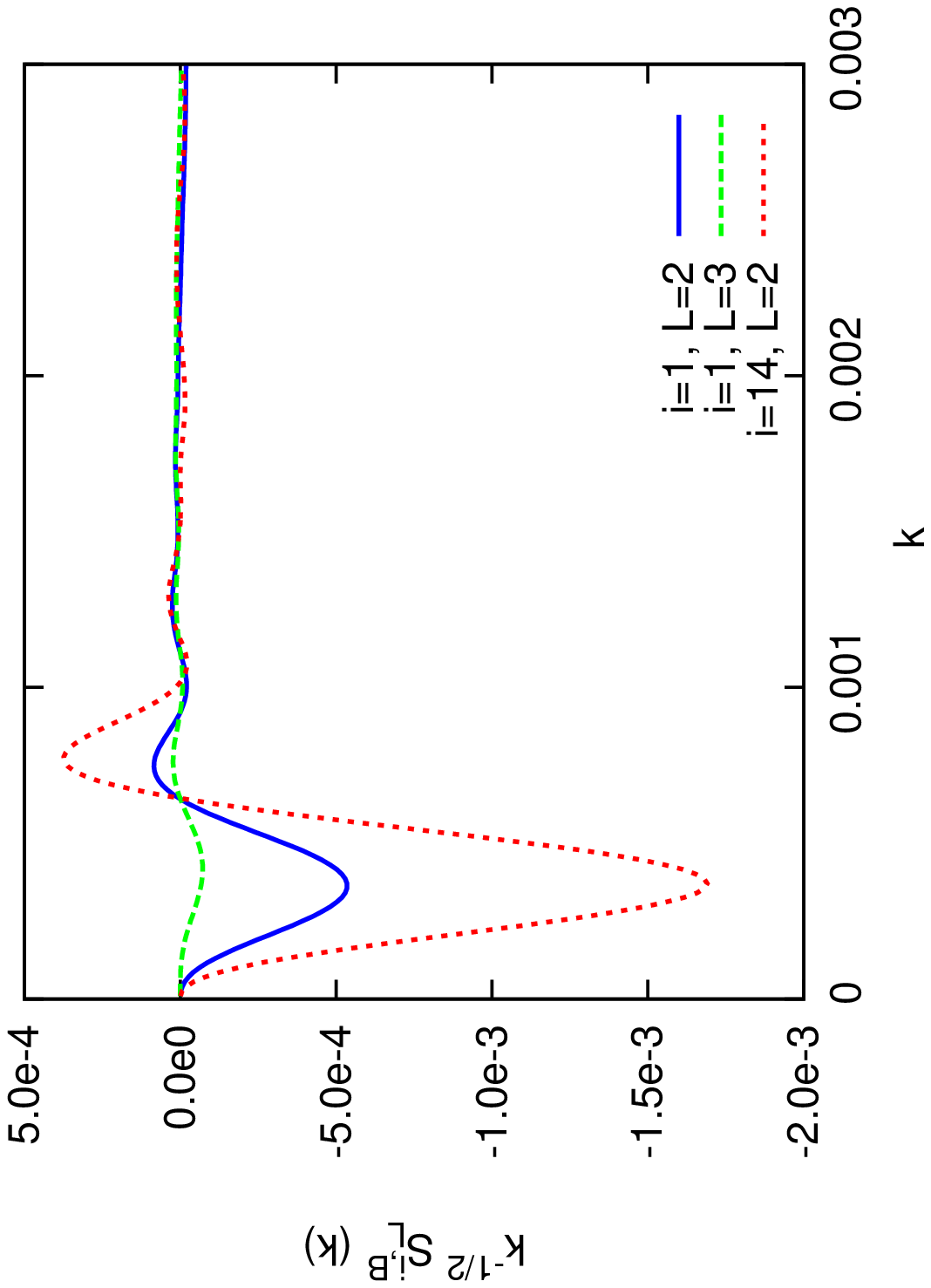}
\caption{[Left] The evolution of the temperature quadrupole generated by the gravitational waves. [Right] Transfer function $S^{i,B}_L$ vs wavenumber $k$, shown for different choices of multipoles $L$ and of redshift bins ``i". }
\label{fig:Ix_Sk}
\end{figure*}

In this section we calculate the signal covariance matrix $C^{\bar{E}^i\bar{E}^j}_l$. We start from the
 line of sight solution to the Boltzmann equation for $E$-type polarization, see e.g.~\cite{1997PhRvD..56..596H},
\begin{equation}
	\bar{E}^i_{lm}(k) \equiv \bar{E}^i_{lm}(\eta=\eta_0,k) =
		 -\sqrt{6}(2l+1) \int_{\eta_i}^{\eta_i+\Delta \eta_i} d\eta \bar{g}(\eta) P_m(\eta,k) \epsilon_{lm}(k(\eta_0-\eta)) .
\label{equ:B1}		 
\end{equation}
In coordinates where $\hat{\mathbf{z}} \parallel \mathbf{k}$, $m=0,1,2$ correspond to the scalar, vector, and tensor modes, respectively.
The projection functions for scalar and tensor modes are
\begin{eqnarray}
	\epsilon_{l,0}(x) &=& \sqrt{\frac{3}{8} \frac{(l+2)!}{(l-2)!}} \frac{j_l(x)}{x^2} , \nonumber  \\
	\epsilon_{l,2}(x) &=& \frac{1}{4}\left[-j_l(x) + j''_l(x)+\frac{2j_l(x)}{x^2} + \frac{4j' _l(x)}{x}\right], 
\end{eqnarray}
and the source functions are
\begin{eqnarray}
	P_m = \frac{1}{10}\left[ \Theta_{2,m} - \sqrt{6} E_{2,m} \right] \simeq \frac{1}{10} \Theta_{2,m} ,
\label{equ:B4}	
\end{eqnarray}
where we have ignored the polarization term in the source equation above compared to the temperature term. The temperature multipoles 
themselves can also be found as line of sight integrals
\begin{eqnarray}
	\frac{\Theta_{l,0}(\eta,k)}{2l+1} &=& \int_0^\eta d\eta' e^{-\tau(\eta,\eta')} \left[ (\dot{\tau} \Theta_{0,0}
			+\dot{\tau} \Psi + \dot{\Psi} -\dot{\Phi})j_{l,0,0} + \dot{\tau} v^B_0 j_{l,1,0} + \dot{\tau} P_0 j_{l,2,0}\right] ,
			\nonumber \\
	\frac{\Theta_{l,2}(\eta,k)}{2l+1} &=& \int_0^\eta d\eta' e^{-\tau} \left[ \dot{\tau} P_2(\eta',k) - \dot{H}^{(+2)}(\eta',k)\right]
			j_{l,2,2}(k(\eta-\eta') ,
\end{eqnarray}
for the scalar and tensor modes, respectively. Here the optical depth is $\tau(\eta,\eta') = \int_{\eta'}^{\eta} \dot{\tau}(\eta'')d\eta''$ 
and $H^{(+2)}$ is the amplitude of the right handed gravitational wave. The projection functions are
\begin{eqnarray}
	j_{l,0,0}(x) &=& j_l(x), \;\;\; j_{l,1,0}=j'_l(x), \; \;\; j_{l,2,0}(x) = \frac{1}{2} \left[3j''_l(x) + j_l(x)\right] , 
	\nonumber \\
	j_{l,2,2}(x) &=& \sqrt{\frac{3}{8} \frac{(l+2)!}{(l-2)!}} \frac{j_l(x)}{x^2} .
\end{eqnarray}
Assuming that recombination occurred instantaneously at $\eta=\eta_*$, neglecting the scattering source term $\dot{\tau}P_2$ compared
to redshifting $\dot{H}$, and writing the metric perturbation in terms of the intial metric perturbation and the transfer function $H^{(+2)}(\eta,k)=T(\eta k) H^{(+2)}_{in}(k)$,
 we find for the temperature quadrupole coming from right-handed gravitational waves
\begin{eqnarray}
	\Theta_{2,2}(\eta,k) = -5 H^{(+2)}_{in}(k)\left[I(k\eta)-I(k\eta_*)\right] , \label{equ:B9} \\
	I(x) \equiv \int^x dx' \frac{T(x')}{dx'} j_{2,2,2}(x-x') .
\end{eqnarray}
The transfer function in the matter dominated universe is
\begin{equation}
	T(x) = \frac{3j_1(x)}{x} .
\end{equation}
We show a plot of $I(x)$ in Fig.~\ref{fig:Ix_Sk}.
To find $C^{\bar{E}^i \bar{E}^j}_{l}$ we notice that the multipoles $\bar{E}^i_{lm}(k)$ in equation \ref{equ:B1} are different from the 
$\bar{E}^i_{lm}(k)$ we used in our calculation in the main text because instead of $\Y{\pm2}{}{l}{m}{n}$ they are expanded in terms of
\begin{equation}
 (-i)^l \sqrt{\frac{4 \pi}{2l+1}} \Y{\pm2}{}{l}{m}{n} .
\end{equation}
Correcting for this normalization and assuming equal amounts of left and right circularly polarized waves, 
the power spectrum of the tensor modes can be found as
\begin{equation}
	C^{\bar{E}^i \bar{E}^j}_{l,T} = \frac{4}{\pi (2l+1)^2} \int dk k^2 P^{\bar{E}^i \bar{E}^j}_{l,2}(k) ,
\label{equ:B12}	
\end{equation}
where $P^{\bar{E}^i \bar{E}^j}_{l,2}(k)$ is defined as
\begin{equation}
	\langle \bar{E}^{i*}_{l,2}(k) \bar{E}^j_{l,2}(k')\rangle = (2\pi)^3 P^{\bar{E}^i \bar{E}^j}_{l,2}(k) \delta^3(k-k') .
\label{equ:B13}	
\end{equation}

Using equations \ref{equ:B1}, \ref{equ:B4}, \ref{equ:B9}, \ref{equ:B12} and \ref{equ:B13} one finally finds

\begin{eqnarray}
	C^{\bar{E}^i \bar{E}^j}_{l,T} = \frac{4}{\pi} \int dk k^2 S^{i,E}_l(k) S^{j,E}_l(k) P_{H^{(+2)}_{in}}(k), \nonumber \\
	S^{i,E}_l(k) = \frac{\sqrt{6}}{2} \int_i d\eta \bar{g}(\eta) \epsilon_{l,2}(k(\eta_0-\eta))\left(I(k\eta)-I(k\eta_*)\right),\label{equ:App_CEiEi}
\end{eqnarray}
where $P_{H^{(+2)}_{in}}$ is the power spectrum of the initial right handed GWs. To make a connection to the definition of $P_h$ in 
reference~\cite{2009ApJS..180..330K} we notice that $H^{(+2)} = -\frac{1}{2}\sqrt{\frac{2}{3}} (\tilde{h}_+ - i \tilde{h}_{\times})$ and therefore
\begin{equation}
	 P_{H^{(+2)}_{in}}(k) = \frac{1}{6} (\langle |\tilde{h}_+|^2 \rangle + \langle |\tilde{h}_{\times}|^2 \rangle ) 
	 									 = \frac{1}{3} \langle |\tilde{h}|^2 \rangle = \frac{P_h(k)}{12}  .
\end{equation}
 
An equation similar to Eq.~\ref{equ:App_CEiEi} can be found for $C^{\bar{B}^i \bar{B}^j}_{l,T}$
 by substituting $\epsilon_{l,2} \rightarrow \beta_{l,2}$ where
\begin{equation}
	\beta_{l,2}(x) = \frac{1}{2} \left[j'_l(x) + 2\frac{j_l(x)}{x} \right] .
\end{equation}	 
In the right panel of Fig.~\ref{fig:Ix_Sk} we show the transfer function S(k) for the quadrupole moment of the $\bar{B}$ polarization 
generated at the first redshift bin.

\section{Sky Cuts}
\label{app:SCu}
 Here we present the detailed calculation the degradation in our estimation for the tensor-to-scalar ratio caused by the sky cuts. As mentioned before, these cuts are troublesome for a multipole-space based estimator because $E$ and $B$ modes cannot be found unambiguously in this case. Fortunately, the reconstruction of the mean polarization field in the $i^{\rm th}$ redshift slice, $\bar P^i(\hat{\bf n})$, is a ``local'' operation in the sense that it only depends on information within a few degrees of $\hat{\bf n}$ (see below and Ref.~\cite{2004PhRvD..70j3501H}). However, this still leaves us with the problem of computing the large-scale power spectrum $C_\ell^{BB\,ij}$ and forecasting its errors in the presence of a cut sky. This is the same problem that occurs in the context of the CMB power spectrum and many works have been devoted to the problem~\citep{1997MNRAS.289..285H,1997MNRAS.289..295H,1997PhRvD..55.5895T,2003NewA....8..581P}.

If we define the filtered fields 
 \begin{eqnarray}
	&&\blsub{\pm} P^F(\nhat) \equiv \sum_{l_1 m_1} \left( \frac{E^{\rm obs}_{l_1 m_1}}{C^{EE,{\rm obs}}_{l_1}} \pm i \frac{B^{\rm obs}_{l_1 m_1}}{C^{BB,{\rm obs}}_{l_1}} \right) \Y{\pm 2}{}{l_1}{m_1}{n} ,
\nonumber \\
	&&\Delta^{F,i}(\nhat) \equiv 	\sum_{l_2 m_2} \frac{C^{g^i g^i}_{l_2} }{b_i C^{g^i g^i,{\rm obs}}_{l_2}} \Delta^{i,{\rm obs}}_{l_2 m_2}
		\Y{}{}{l_2}{m_2}{n}, 
\end{eqnarray}
 then one can prove that
 \begin{eqnarray}
 	 \blsub{\pm} \hat{\bar{P}}^{F,i} (\nhat) \equiv \blsub{\pm} P^F(\nhat) \Delta^{F,i}(\nhat) = \sum_{LM}   \frac{ \hat{\bar{E}}^i_{LM} \pm i \hat{\bar{B}}^i_{LM} }{ A^{i,B}_L }     \Y{\pm 2}{}{L}{M}{n} .
 \end{eqnarray}
  We have then accomplished our goal of constructing an estimator for the average polarization generated at a given redshift bin that is not affected by the eliminated portions of the sky, since the quantities $\blsub{\pm} P^F(\nhat)$ and $\Delta^{F,i}(\nhat)$ are local in configuration space. We can further simplify the analysis by noting that the $A^{i,B}_L$ are very nearly independent of scale for small $L$, where most of the signal comes from. Thus we can pull this factor out of the sum and find an estimator for the non-filtered average polarization generated at the redshift bin ``i"
\begin{eqnarray}
	 	\blsub{\pm}\hat{\bar{P}}^{i} (\nhat) \equiv \sum_{LM} \left( \hat{\bar{E}}^i_{LM} \pm i \hat{\bar{B}}^i_{LM} \right) \Y{\pm 2}{}{L}{M}{n} = A^{i,B}_2 \blsub{\pm} \hat{\bar{P}}^{F,i} (\nhat) .
\end{eqnarray}
 The estimators for the Stokes parameters, $\hat{\bar{Q}}^i(\nhat)$ and $\hat{\bar{U}}^i(\nhat)$, can then be found as
\begin{eqnarray}
	\hat{\bar{Q}}^i(\nhat) &=& \frac{\blsub{+}\hat{\bar{P}}^i (\nhat) + \blsub{-}\hat{\bar{P}}^i (\nhat)}{2} 
			=\sum_{LM} \hat{\bar{E}}^i_{LM} Y^E_{LM}(\nhat,Q) + \hat{\bar{B}}^i_{LM} Y^B_{LM}(\nhat,Q) , \nonumber \\
	\hat{\bar{U}}^i(\nhat) &=& \frac{\blsub{+}\hat{\bar{P}}^i (\nhat) - \blsub{-}\hat{\bar{P}}^i (\nhat)}{2 i} 
			=\sum_{LM} \hat{\bar{E}}^i_{LM} Y^E_{LM}(\nhat,U) + \hat{\bar{B}}^i_{LM} Y^B_{LM}(\nhat,U) ,
\end{eqnarray} 
 where we have defined
\begin{eqnarray}
 Y^E_{LM}(\nhat,Q) &=& Y^B_{LM}(\nhat,U) = \frac{1}{2} \left ( \Y{2}{}{L}{M}{n} + \Y{-2}{}{L}{M}{n} \right) , \nonumber \\
 Y^B_{LM}(\nhat,Q) &=& -Y^E_{LM}(\nhat,U) = \frac{i}{2} \left ( \Y{2}{}{L}{M}{n} - \Y{-2}{}{L}{M}{n} \right) .
 \label{eq:tensorY}
\end{eqnarray}
We write down a vector ${\bf x}$ consisting of the mean polarization $\bar Q^i(\hat{\bf n}_A)$ and $\bar U^i(\hat{\bf n}_A)$ in the $A^{\rm th}$ pixel, of length $2N_zN_{\rm pix}$. Then ${\bf x}$ has a $2N_zN_{\rm pix}\times 2N_zN_{\rm pix}$ covariance matrix
\begin{equation}
C^{ij}_{\chi_A\chi_B}(\hat{\bf n}_A,\hat{\bf n}_B),
\label{eq:cij}
\end{equation}
where $\chi_A$ indicates a choice of Stokes parameter ($Q$ or $U$).
This contains a contribution ${\bf C}_0$ arising from all sources other than gravitational waves (including the scalar post-reionization scattering signal and noise), as well as a contribution associated with tensor modes,
\begin{equation}
{\bf C}={\bf C}_0 + r{\bf C}_r.
\end{equation}
If it is desired to remove certain modes from the data (e.g. $E$-modes), then they can be incorporated in ${\bf C}_0$ by introducing such modes with formally infinite power. (This is a common foreground template projection method in CMB data analysis, e.g. \cite{2004PhRvD..69l3003S}.) 

Under the null hypothesis of no primordial gravitational waves, and assuming Gaussian signal and noise, one can write down the Fisher information,
\begin{equation}
F_{rr} = \frac12{\rm Tr}\,\left( {\bf C}_0^{-1}{\bf C}_r{\bf C}_0^{-1}{\bf C}_r \right),
\end{equation}
and forecast an uncertainty $\sigma_r = F_{rr}^{-1/2}$. The assumption of Gaussian signal and noise is expected to be valid in this case: the signal (primordial gravitational waves) are Gaussian and feed linearly into $\bar P^i(\hat{\bf n})$. The noise (associated with the product of primary CMB and large scale structure) is not Gaussian, but its average in a large patch of sky (tens of degrees) contains contributions from many arcminute-scale patches, so we expect the noise contribution to $\bar P^i(\hat{\bf n})$ to be Gaussianized by the central limit theorem.

The actual computation of $F_{rr}$ is simplified if we consider only the $B$-mode quadrupole, $\ell = 2$, and note that the noise power spectrum is white at low $\ell$ (i.e. it becomes independent of $\ell$ and equal for $E$ and $B$ modes). In this case, the matrices have the form
\begin{equation}
[{\bf C}_r]^{ij}_{\chi_A\chi_B}(\hat{\bf n}_A,\hat{\bf n}_B) = C_2^{\bar B^i\bar B^j} \sum_{m=-2}^2
Y_{2m}^{B\ast}(\hat{\bf n}_A,\chi_A)
Y_{2m}^{B}(\hat{\bf n}_B,\chi_B),
\end{equation}
where $Y_{2m}^B(\hat{\bf n},\chi)$ denotes the $\chi$ component of the $B$-mode tensor spherical harmonic \cite{1997PhRvD..55.7368K} evaluated at $\hat{\bf n}$ (see Eq.~\ref{eq:tensorY}).
If we insist on projecting out all $E$-mode signals (so that there is no possible contamination from the scalar reionization signal), then the zero-tensor covariance matrix is
\begin{eqnarray}
[{\bf C}_0]^{ij}_{\chi_A\chi_B}(\hat{\bf n}_A,\hat{\bf n}_B)
\!\!&=&\!\!
N^{ij}\delta_{\chi_A\chi_B}\delta^{(2)}(\hat{\bf n}_A-\hat{\bf n}_B)
\nonumber \\
&& + \sum_{\ell=2}^{\ell_{\rm max}} \lambda_\ell \delta^{ij} \sum_{m=-\ell}^\ell
Y_{2m}^{E\ast}(\hat{\bf n}_A,\chi_A)
Y_{2m}^{E}(\hat{\bf n}_B,\chi_B),
\end{eqnarray}
where $N^{ij}$ is the white noise level (still a matrix because it depends on the redshift slices), and the second term projects out $E$-mode signals. The projection parameters $\lambda_\ell$ should formally be taken to $\infty$.

Under the above assumptions, the matrix inversions and trace factor into pieces that depend on the sky coverage, and pieces that depend on the signal and noise power spectra ($C_2^{\bar B^i\bar B^j}$ and $N^{ij}$):
\begin{equation}
F_{rr} = \frac12 {\rm Tr}\left( {\bf N}^{-1}{\bf C}_2^{\bar B\bar B}{\bf N}^{-1}{\bf C}_2^{\bar B\bar B}\right)
 {\rm Tr}\left( {\bf K}^{-1} {\bf S}{\bf K}^{-1}{\bf S} \right),
\label{eq:frr}
\end{equation}
where the first trace is of a product of $N_z\times N_z$ matrices, and the second trace is of a product of the $2N_{\rm pix}\times 2N_{\rm pix}$ matrices:
\begin{equation}
K_{\chi_A\chi_B}(\hat{\bf n}_A,\hat{\bf n}_B) = \delta_{\chi_A\chi_B}\delta^{(2)}(\hat{\bf n}_A-\hat{\bf n}_B) + \sum_{\ell=2}^{\ell_{\rm max}} \lambda_\ell \sum_{m=-\ell}^\ell
Y_{2m}^{E\ast}(\hat{\bf n}_A,\chi_A)
Y_{2m}^{E}(\hat{\bf n}_B,\chi_B)
\end{equation}
(again, valid only in the $\lambda_{\ell}\rightarrow\infty$ limit) and
\begin{equation}
S_{\chi_A\chi_B}(\hat{\bf n}_A,\hat{\bf n}_B) = \sum_{m=-2}^2
Y_{2m}^{B\ast}(\hat{\bf n}_A,\chi_A)
Y_{2m}^{B}(\hat{\bf n}_B,\chi_B).
\end{equation}

In Eq.~(\ref{eq:frr}), the first trace depends on detailed galaxy and CMB properties and has been computed in the main paper. The second trace enables us to define a ``degradation factor:''
\begin{equation}
f_{\rm deg} \equiv \frac{
{\rm Tr}\left( {\bf K}^{-1} {\bf S}{\bf K}^{-1}{\bf S} \right)({\rm true})
}{
{\rm Tr}\left( {\bf K}^{-1} {\bf S}{\bf K}^{-1}{\bf S} \right)({\rm all~sky})
}.
\label{eq:fdeg}
\end{equation}
It is $f_{\rm deg}$ that this appendix aims to compute. For a cut sky, the uncertainty on $r$ increases by a factor of $f_{\rm deg}^{-1/2}$. The analytic expectation is that in the harmonic basis and on the full sky, ${\bf K}$ becomes a diagonal matrix with 1s in the $B$-mode positions and $\infty$s in the $E$-mode positions (because of the formally infinite $E$-mode terms added); ${\bf S}$ should have 1s in the diagonal $\ell=2$ $B$-mode positions and 0s elsewhere. Thus the denominator should simply be the number of $\ell=2$ $B$-modes, i.e. 5. We find that numerically this is indeed the case.

\bibliographystyle{h-physrev}
\bibliography{ref}
\end{document}